\documentclass[a4paper,11pt]{article}
\pdfoutput=1
\usepackage{jcappub}
\usepackage{multirow}
\usepackage{comment}
\usepackage{booktabs}
\usepackage{xspace}
\usepackage{subcaption}
\usepackage[dvipsnames]{xcolor}

\newcommand{\meft}{mEFT\xspace}
\newcommand{\lya}{Ly$\alpha$\xspace}
\newcommand{\hMpc}{\,h\,{\rm Mpc}^{-1}}
\newcommand{\Mpch}{\,h^{-1}{\rm Mpc}}
\newcommand{\kmax}{k_{\rm max}}
\newcommand{\Tb}{\bar T_{\rm b}}
\newcommand{\bHI}{b_{\rm HI}}
\newcommand{\kt}{\tilde k}

\newcommand{\lref}[2]{\hyperref[#2]{\mbox{#1~\ref*{#2}}}}
\newcommand{\leqref}[2]{\hyperref[#2]{\mbox{#1\,(\ref*{#2})}}}

\title{The 21\,cm--Lyman-$\alpha$ forest cross-power spectrum with a minimal effective field theory}

\author[a,b,c]{Gabriele Autieri,}
\author[d]{Vid Ir\v{s}i\v{c},}
\author[a,b,c,e,f]{Matteo Viel}

\affiliation[a]{SISSA -- International School for Advanced Studies, Via Bonomea 265, 34136 Trieste, Italy}
\affiliation[b]{INFN, Sezione di Trieste, Via Valerio 2, 34127 Trieste, Italy}
\affiliation[c]{IFPU -- Institute for Fundamental Physics of the Universe, Via Beirut 2, 34151 Trieste, Italy}
\affiliation[d]{Centre for Astrophysics Research, Department of Physics, Astronomy and Mathematics, University of Hertfordshire, Hatfield, AL109AB, UK}
\affiliation[e]{INAF, Osservatorio Astronomico di Trieste, Via G. B. Tiepolo 11, I-34131 Trieste, Italy}
\affiliation[f]{ICSC - Italian Research Center on High Performance Computing, Big Data and Quantum Computing, via Magnanelli 2, 40033, Casalecchio di Reno, Italy}

\emailAdd{gautieri@sissa.it}

\abstract{
Intensity maps of the 21\,cm line and the Lyman-$\alpha$ (\lya) forest trace the same cosmic volume at $2\lesssim z\lesssim 4$. Their cross-power spectrum is insensitive to foregrounds that do not correlate between the two probes and, together with the \lya auto-spectrum, separates the mean 21\,cm brightness temperature from the neutral hydrogen (HI) bias. Effective field theory models of each tracer exist, but at one loop they carry tens of nuisance parameters. We test whether a minimal effective field theory (\meft), consisting of the tree-level Kaiser model with leading-order counterterms and stochastic terms, previously validated for the \lya auto-spectrum, describes the 21\,cm auto-, \lya auto- and cross-power spectra jointly. We use the Sherwood 80\,$h^{-1}$Mpc hydrodynamical simulation with HI painted onto halos at $z=2.4$--$4.2$, and reduce sample variance with Zel'dovich control variates. Each spectrum is validated separately and then in a joint fit with shared parameters. With scale cuts $\kmax=0.5$, $1.2$ and $2.0\,h\,{\rm Mpc}^{-1}$ for the 21\,cm auto-, cross- and \lya auto-spectra, the \meft fits all three spectra at all five redshifts. In the joint fit the \meft determines the HI bias to 2.0--2.5\% and the \lya density bias to 2.1--2.3\%. The scale cuts follow the HI velocity dispersion and the \lya pressure-smoothing scale. When the brightness temperature is left free, the fits that include the cross-spectrum prefer values 20--30\% above the true one, at 1--2$\sigma$, while the two auto-spectra together do not constrain it. A prior on the HI bias recovers the true value of the brightness temperature.
}

\begin{document}
\maketitle
\flushbottom

\section{Introduction}
\label{sec:intro}

After reionisation, most of the neutral hydrogen (HI) in the Universe resides in dense, self-shielded systems inside dark matter halos, and the integrated 21\,cm emission of many unresolved galaxies traces the large-scale matter distribution~\cite{Bharadwaj:2001,Wyithe:2008,Loeb:2008,Chang:2008,Furlanetto:2006,Pritchard:2012}. Line-intensity mapping of the 21\,cm line measures the integrated emission over large volumes without resolving individual sources, so large areas of the sky can be surveyed quickly and at relatively low cost~\cite{Kovetz:2017,Bull:2015}. The signal has been detected in cross-correlation with galaxy surveys at $z\lesssim1.4$~\cite{Chang:2010,Masui:2013,Wolz:2022,Amiri:2023,Cunnington:2022, 2025MNRAS.537.3632M}, in auto-correlation on small and large scales at low redshift~\cite{Switzer:2013,Paul:2026, MeerKLASS:2026ztn}, and at $z\simeq2$--$2.5$ in cross-correlation with the Lyman-$\alpha$ (\lya) forest measured by eBOSS~\cite{Amiri:2024}. Current and planned interferometers such as CHIME~\cite{Amiri:2022}, HIRAX~\cite{Newburgh:2016,Crichton:2022}, CHORD~\cite{Vanderlinde:2019} and a PUMA-like array~\cite{Slosar:2019,Ansari:2018}, as well as single-dish surveys with MeerKAT and the SKA~\cite{Santos:2015,Santos:2017,Bacon:2020}, aim to map HI over a wide redshift range, including $z>2$ where the matter field is closer to linear and the volume is large~\cite{Obuljen:2018,Castorina:2019,Sailer:2021}. The main difficulty is intrinsic to the measurement: Galactic and extragalactic foregrounds are orders of magnitude brighter than the signal and remove the modes with small line-of-sight wavenumber~\cite{Liu:2020}.

The \lya forest in the spectra of distant quasars traces the neutral fraction of the diffuse intergalactic medium at $2\lesssim z\lesssim5$~\cite{Meiksin:2009,McQuinn:2016}. Its three-dimensional clustering, first measured with BOSS~\cite{Slosar:2011}, now provides baryon acoustic oscillation measurements at the percent level~\cite{duMasdesBourboux:2020,Adame:2025,AbdulKarim:2025}, and analyses of the full shape of its correlation functions and power spectrum have started~\cite{Cuceu:2023,Cuceu:2025,Adame:2026,FontRibera:2018,Horowitz:2024}. On large scales the \lya flux contrast is a biased tracer of the matter field with a density bias $b_{\rm F}<0$ and a velocity-gradient bias $b_\eta$~\cite{McDonald:2003,Seljak:2012,Cieplak:2016}. On small scales the \lya forest is shaped by the thermal state and pressure smoothing of the intergalactic gas, which are difficult to model from first principles. For this reason, its non-linear power spectrum has usually been described with fitting functions calibrated on hydrodynamical simulations~\cite{McDonald:2003,ArinyoiPrats:2015,Givans:2022,Soltinsky:2026rbi}.

Because both probes trace the same volume at $2\lesssim z\lesssim4$, their cross-correlation has been proposed as a way to measure the HI clustering and cosmology with a signal that is insensitive to foregrounds and systematics that do not correlate between radio maps and optical spectra~\cite{GuhaSarkar:2011,GuhaSarkar:2013,Sarkar:2018,Sarkar:2019,MonteroCamacho:2024,Pal:2026}. Together with the \lya auto-spectrum, the cross-spectrum also separates $\Tb$ from $\bHI$. Simulation-based studies of the cross-power spectrum~\cite{Carucci:2017} and of 21\,cm maps correlated with other high-redshift tracers~\cite{VillaescusaNavarro:2015} have so far relied on linear theory or fitting functions, and the forecasts cited above have used linear or phenomenological models.

The effective field theory of large-scale structure (EFTofLSS)~\cite{Baumann:2012,Carrasco:2012,Senatore:2014,Perko:2016,Desjacques:2018} provides a systematic perturbative description of biased tracers in redshift space, and its one-loop predictions are the standard model for full-shape analyses of galaxy clustering~\cite{Ivanov:2020,DAmico:2020,Chudaykin:2020}. It has been applied to HI in simulations and forecasts~\cite{Sailer:2021,Obuljen:2023,Baradaran:2024} and, more recently, to the \lya forest~\cite{Ivanov:2024,Givans:2022,deBelsunce:2025,Chudaykin:2025,deBelsunce:2026,deBelsunce:2026b}, including a cosmological analysis of the one-dimensional flux power spectrum~\cite{Ivanov:2025}. At one loop, the \lya forest requires 11 bias parameters in addition to counterterms and stochastic terms~\cite{Ivanov:2024}. Constraints on these parameters then depend on their priors, which has motivated priors calibrated on simulations~\cite{Ivanov:2025,Ivanov:2024b,Ivanov:2025c,Zhang:2025,Ivanov:2025b,Sarkar:2026} and studies of projection effects in EFTofLSS analyses~\cite{Simon:2023}.

In ref.~\cite{Autieri:2026} we showed that, for the three-dimensional \lya forest in the Sherwood simulations~\cite{Bolton:2017,Puchwein:2023}, a minimal effective field theory consisting of the tree-level Kaiser model with the leading counterterms and stochastic terms, which we call \meft, describes the flux auto-power spectrum and its cross-spectrum with the matter field up to $\kmax=2$--$3\hMpc$. Here we ask whether the same minimal model holds for the 21\,cm auto-spectrum and for the 21\,cm--\lya cross-spectrum. The question is non-trivial for two reasons. First, HI sits in halos whose internal velocities produce fingers of God on scales that are not much smaller than those of interest. Second, in a joint fit of the three spectra with shared bias parameters, sample variance largely cancels~\cite{Seljak:2009,McDonald:2009b}, so the joint fit is a more stringent test of the model than any single spectrum.

We use the Sherwood $80\Mpch$ simulation at $z=2.4$--$4.2$, paint HI onto its halos following ref.~\cite{VillaescusaNavarro:2018}, and reduce the sample variance of the measured spectra with Zel'dovich control variates~\cite{Kokron:2022,DeRose:2023}. We fit the 21\,cm auto-, \lya auto- and cross-power spectrum multipoles with the \meft, first one spectrum at a time and then jointly, and examine the scale cuts, the linear biases, the consistency between data combinations and the recovery of $\Tb$. The comparison of the \meft with the one-loop effective field theory of biased tracers on the same data is left for future work.

The paper is organised as follows. \lref{Section}{sec:data} describes the simulation, the HI painting, the power spectrum measurements, the control variates and the covariance. \lref{Section}{sec:theory} presents the model and the relevant physical scales. \lref{Section}{sec:cuts} summarises the scale cuts, which are validated in \lref{appendix}{app:cuts}. \lref{Section}{sec:meft} gives the \meft results. We conclude in \lref{Section}{sec:conclusions}. 
\section{Simulations and data}
\label{sec:data}

\subsection{The Sherwood simulation and the \texorpdfstring{\lya}{Lyα} forest}
\label{sec:sherwood}

We use the $80\Mpch$ box with $2\times1024^3$ gas and dark matter particles of the Sherwood simulation suite~\cite{Bolton:2017}, run with a modified version of \textsc{P-Gadget-3}~\cite{Springel:2005}. The cosmology is flat $\Lambda$CDM with $\Omega_{\rm m}=0.308$, $\Omega_{\rm b}=0.0482$, $h=0.678$, $\sigma_8=0.829$ and $n_{\rm s}=0.961$~\cite{Ade:2014}. We analyse the snapshots at $z=2.4$, $2.8$, $3.2$, $3.6$ and $4.2$.

The \lya forest flux is computed as in ref.~\cite{Autieri:2026} (see ref.~\cite{Soltinsky:2026rbi} for further details). The optical depth is extracted along $1024^2$ sightlines parallel to one axis of the box, each sampled with 1024 pixels, so that the transmitted flux $F=e^{-\tau}$ is defined in redshift space on a regular $1024^3$ grid. The optical depths are rescaled by a constant at each redshift to match the observed mean transmitted flux of ref.~\cite{Viel:2013}, as in ref.~\cite{Autieri:2026}.
We use the flux contrast $\delta_{\rm F}=F/\bar F-1$.

\subsection{HI painting}
\label{sec:painting}

We assign HI to the friends-of-friends (FoF) halos of each snapshot, without any mass threshold beyond the resolution of the halo catalogue. Each halo of mass $M$ receives an HI mass defined by the form calibrated on the IllustrisTNG simulations in ref.~\cite{VillaescusaNavarro:2018}
\begin{equation}
M_{\rm HI}(M,z)=M_0(z)\left(\frac{M}{M_{\rm min}(z)}\right)^{\alpha(z)}\exp\!\left[-\left(\frac{M_{\rm min}(z)}{M}\right)^{0.35}\right].
\label{eq:mhi}
\end{equation}
We take the FoF parameters of that work, tabulated at $z=2,3,4,5$, and interpolate them to our redshifts linearly in $\alpha$ and in $\log M_0$ and $\log M_{\rm min}$. The values are listed in \lref{Table}{tab:sim}. At $z=2.4$ we use the rounded values $M_0=7.13\times10^9\,h^{-1}M_\odot$, $\alpha=0.664$ and $M_{\rm min}=1.84\times10^{11}\,h^{-1}M_\odot$.
The HI of each halo is placed at the halo centre with a cloud-in-cell (CIC) assignment on a $1024^3$ grid, and the field is mapped to redshift space along the same axis as the \lya sightlines. The halo positions are displaced by the halo peculiar velocities. To include the internal motions of HI within halos, the HI of each halo is in addition spread along the line of sight with a Gaussian profile of comoving width~\cite{Sarkar:2019nak, Zhang:2019fvu}
\begin{equation}
\sigma_{\rm s}(M,z)=\frac{(1+z)\,\sigma_{\rm v}(M,z)}{\sqrt3\,H(z)},\qquad \sigma_{\rm v}(M,z)=\sigma_{10}(z)\left(\frac{M}{10^{10}\,h^{-1}M_\odot}\right)^{\alpha_{\rm v}(z)},
\label{eq:fog}
\end{equation}
where $\sigma_{\rm v}$ is the three-dimensional HI velocity dispersion of a halo of mass $M$ measured in IllustrisTNG, fitted with $\sigma_{10}(z)=(4.89\,z+29.95)\,{\rm km\,s^{-1}}$ and $\alpha_{\rm v}(z)=0.0091\,z+0.357$. For halos of $10^{11}$--$10^{12}\,h^{-1}M_\odot$, typical HI hosts at these redshifts, \leqref{Eq.}{eq:fog} gives $\sigma_{\rm s}\simeq0.55$--$1.3\Mpch$ at $z=2.4$, i.e.\ fingers of God on scales $k\sim1/\sigma_{\rm s}\simeq0.8$--$1.8\hMpc$ (\lref{Table}{tab:scales}).

The 21\,cm brightness-temperature fluctuation is $\delta T_{\rm b}=\Tb\,\delta_{\rm HI}$, with~\cite{VillaescusaNavarro:2018}
\begin{equation}
\Tb(z)=189\,h\,\frac{(1+z)^2}{E(z)}\,\Omega_{\rm HI}(z)\ {\rm mK},
\label{eq:tb}
\end{equation}
where $E(z)=H(z)/H_0$ and $\Omega_{\rm HI}$ is the total painted HI mass divided by the critical density and the box volume. The painted $\Omega_{\rm HI}=(7.2$--$8.7)\times10^{-4}$ (\lref{Table}{tab:sim}) is somewhat below the values of about $10^{-3}$ inferred from damped \lya systems at these redshifts (e.g.\ ref.~\cite{Crighton:2015}). Because $\Tb$ is known exactly in the simulation, we fix it to its true value in all fits except in \lref{Section}{sec:tb}, where we test whether the data can recover it.

The painting neglects HI outside halos and any dependence of the HI content on halo properties other than mass. We do not add instrumental noise, foregrounds or a telescope beam.

\begin{table}
\centering
\small
\begin{tabular}{ccccccccc}
\toprule
$z$ & $M_0$ & $\alpha$ & $M_{\rm min}$ & $\Omega_{\rm HI}$ & $\Tb$ & $\sigma_{10}$ & $\alpha_{\rm v}$ & $f$ \\
 & [$h^{-1}M_\odot$] & & [$h^{-1}M_\odot$] & [$10^{-4}$] & [mK] & [km\,s$^{-1}$] & & \\
\midrule
2.4 & $7.00\times10^{9}$ & 0.660 & $1.84\times10^{11}$ & 7.20 & 0.298 & 41.7 & 0.379 & 0.9715 \\
2.8 & $3.91\times10^{9}$ & 0.728 & $9.38\times10^{10}$ & 7.35 & 0.324 & 43.6 & 0.383 & 0.9796 \\
3.2 & $2.51\times10^{9}$ & 0.766 & $5.31\times10^{10}$ & 7.71 & 0.359 & 45.6 & 0.386 & 0.9850 \\
3.6 & $1.87\times10^{9}$ & 0.778 & $3.34\times10^{10}$ & 8.31 & 0.407 & 47.5 & 0.390 & 0.9888 \\
4.2 & $1.49\times10^{9}$ & 0.780 & $2.08\times10^{10}$ & 8.72 & 0.455 & 50.5 & 0.396 & 0.9924 \\
\bottomrule
\end{tabular}
\caption{HI painting parameters (\leqref{Eqs.}{eq:mhi} and~\hyperref[eq:fog]{(\ref*{eq:fog})}), painted HI density, mean brightness temperature (\leqref{Eq.}{eq:tb}) and linear growth rate at the five redshifts.}
\label{tab:sim}
\end{table}

\subsection{Power spectrum multipoles}
\label{sec:multipoles}

We measure the auto- and cross-power spectra of $\delta T_{\rm b}$ (HI) and $\delta_{\rm F}$ (F) with \textsc{Pylians}~\cite{Pylians}, and project them onto the Legendre multipoles $\ell=0,2,4$ with respect to the line of sight. We denote the three spectra $P_{\rm HI}$ (21\,cm auto, in mK$^2\,h^{-3}$Mpc$^3$), $P_{\rm F}$ (\lya auto, in $h^{-3}$Mpc$^3$) and $P_{\rm X}$ (cross, in mK\,$h^{-3}$Mpc$^3$).

In the lowest $k$-bins, the number of modes is small and their directions are not uniformly distributed in $\mu$, so the estimated multipoles of a field with a pure monopole acquire spurious $\ell=2,4$ components. For each shell we compute the $3\times3$ matrix $W_{\ell\ell'}=\frac{2\ell+1}{N_k}\sum_{i\in{\rm shell}}\mathcal{L}_\ell(\mu_i)\mathcal{L}_{\ell'}(\mu_i)$, which maps the true multipoles $\ell'\le4$ onto the estimated ones, and apply $W^{-1}$ to the measured multipoles (the ``deconvolved'' data)~\cite{DeRose:2023}. The model is evaluated on the same lattice modes and processed in the same way (\lref{Section}{sec:inference}).

\lref{Figure}{fig:data} shows the monopoles and quadrupoles of the three spectra at $z=2.4$ and $4.2$, with and without the control-variate correction of \lref{Section}{sec:zcv}. \lref{Figure}{fig:rhf}(a) shows the cross-correlation coefficient of the monopoles, $-P_0^{\rm X}/(P_0^{\rm HI}P_0^{\rm F})^{1/2}$. It is $0.96$ on the largest scales at all redshifts and falls to $0.45$ ($z=2.4$) and $0.32$ ($z=4.2$) at $k=1.9\hMpc$. This data-level quantity includes the effect of redshift-space distortions and noise, and it shows that the cross-spectrum contains information that is not a simple combination of the two auto-spectra. \lref{Figure}{fig:rhf}(b) shows the reduction of the uncertainty of the monopoles by the control variates (\lref{Section}{sec:zcv}).

\begin{figure}
\centering
\includegraphics[width=\textwidth]{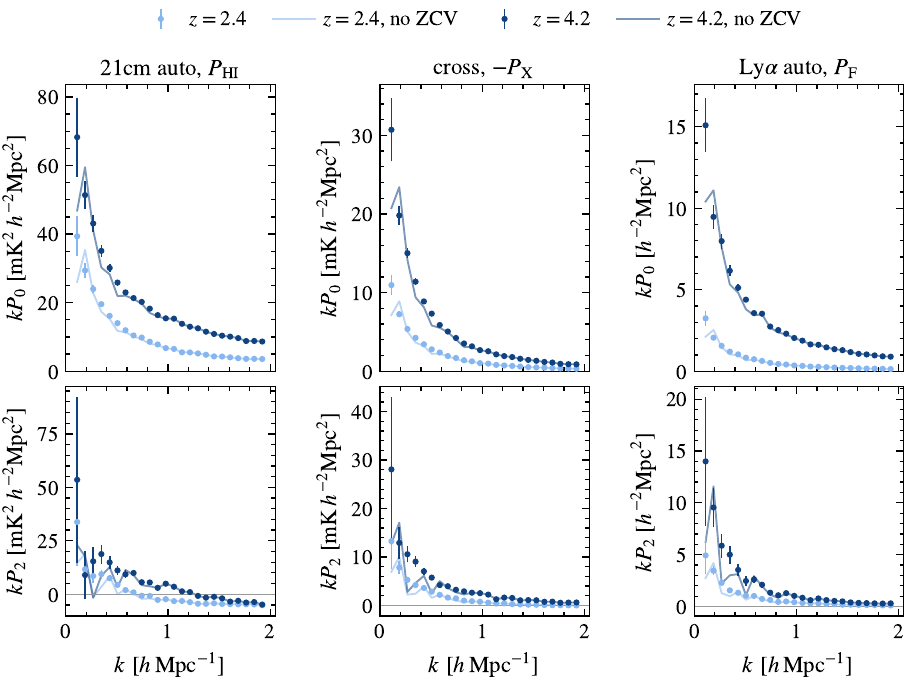}
\caption{Monopole (top) and quadrupole (bottom) of the 21\,cm auto- (left), cross- (centre, sign flipped) and \lya auto-power spectra (right) at $z=2.4$ and $4.2$. The points represent the data after the Zel'dovich control-variate correction, with their uncertainties, whilst the lines denote the data without the correction.}
\label{fig:data}
\end{figure}

\begin{figure}
\centering
\includegraphics[width=\textwidth]{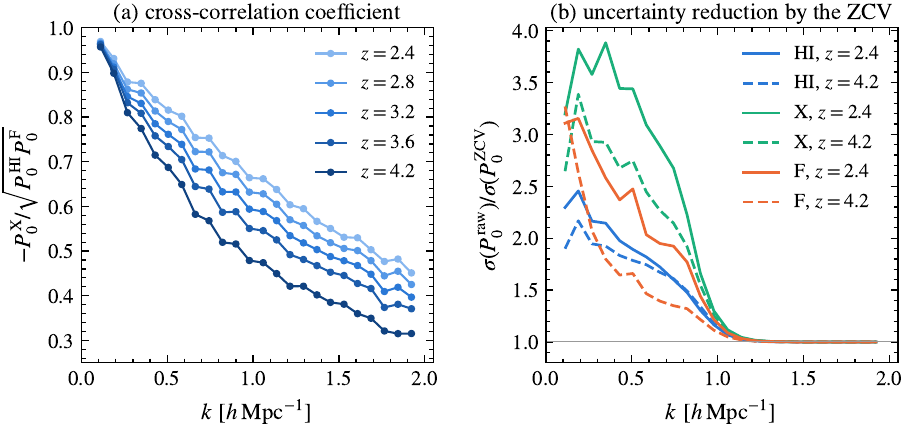}
\caption{(a) Ratio of the cross-power spectrum monopole, with its sign flipped, to the geometric mean of the two auto-spectrum monopoles, at all redshifts. (b) Uncertainty of the monopole without the Zel'dovich control-variate correction relative to that with it, for the three spectra at $z=2.4$ (solid) and $4.2$ (dashed).}
\label{fig:rhf}
\end{figure}

\subsection{Zel'dovich control variates}
\label{sec:zcv}

An $80\Mpch$ box contains few large-scale modes, and the sample variance of the measured multipoles would dominate the uncertainty on the bias parameters. We reduce it with Zel'dovich control variates (ZCV)~\cite{Kokron:2022,DeRose:2023}, following ref.~\cite{Autieri:2026}. For each spectrum $S\in\{P_{\rm HI},P_{\rm X},P_{\rm F}\}$ and multipole we form
\begin{equation}
\hat S_{\rm ZCV}(k)=\hat S(k)-\beta_S(k)\left[\hat Z_S(k)-\langle Z_S\rangle(k)\right],
\label{eq:zcv}
\end{equation}
where $\hat Z_S$ is the corresponding spectrum of a control field computed in the Zel'dovich approximation from the initial conditions of the simulation, $\langle Z_S\rangle$ its ensemble mean and $\beta_S=\mathrm{Cov}(\hat S,\hat Z_S)/\mathrm{Var}(\hat Z_S)$ the coefficient that minimises the variance of $\hat S_{\rm ZCV}$. The control field is generated with \textsc{N-GenIC}~\cite{NGenIC} from the same initial conditions as the simulation, displaced to the target redshift in the Zel'dovich approximation and mapped to redshift space. The coefficients $\beta_S$ are computed from the Gaussian covariance of the measured fields of the same box, i.e.\ from the realisation they correct. As in ref.~\cite{DeRose:2023}, we therefore smooth them in $k$ and damp them smoothly to zero above $k_0=0.9\hMpc$, where the correlation between the control field and the simulated fields is weak. The uncertainty of $\beta_S$ itself is not included in the covariance.

For the ensemble mean we use 50 independent \textsc{N-GenIC} realisations of the control field rather than the analytic Zel'dovich prediction, and its own uncertainty is included in the covariance. At $z=2.4$ and $4.2$ the control variates reduce the uncertainties of the monopoles at $k<0.5\hMpc$ by factors of 1.8--2.5 for the 21\,cm auto-spectrum, 2.6--3.9 for the cross-spectrum and 1.6--3.3 for the \lya auto-spectrum, and by less than 5\% above $k=1.1\hMpc$ (\lref{Figure}{fig:rhf}b).

\subsection{Covariance}
\label{sec:covariance}

We use the hybrid covariance of ref.~\cite{Chudaykin:2025}, adapted to the ZCV data. Below the fiducial scale cuts, $k_{\rm fid}=0.8$, $1.2$ and $2.0\hMpc$ for the 21\,cm auto-, cross- and \lya auto-spectra, the spectra of the simulated fields are replaced by the best-fit model and those of the control field by the ensemble mean, while above $k_{\rm fid}$ the measured modes are kept. The model is the joint best fit of the \meft, except for the 21\,cm auto-spectrum above $0.55\hMpc$, beyond the range of validity of the \meft, where we use the joint best fit of the one-loop EFT of biased tracers~\cite{Ivanov:2020,Ivanov:2024}, obtained with the same covariance. We iterate the construction with the best fits of both models until the joint $\chi^2$ changes by less than 0.3, which takes two to three iterations.
\section{Models}
\label{sec:theory}

We model the redshift-space power spectra of the two tracers, HI ($\delta T_{\rm b}$) and F ($\delta_{\rm F}$), with the \meft. Throughout, $\mu$ is the cosine of the angle between the wavevector and the line of sight, $f$ the linear growth rate, $P_{L}(k)$ the linear matter power spectrum at the redshift of the data computed with \textsc{CAMB}~\cite{Lewis:2000}, and we work at fixed cosmology.

\subsection{The minimal EFT}
\label{sec:meft_model}

On large scales both tracers follow the Kaiser form~\cite{Kaiser:1987} with the kernels
\begin{equation}
K_{\rm HI}(\mu)=\bHI+f\mu^2,\qquad K_{\rm F}(\mu)=b_{\rm F}+b_\eta f\mu^2,
\label{eq:kernels}
\end{equation}
where $b_{\rm F}<0$ and $b_\eta<0$ are the density and velocity-gradient biases of the \lya forest~\cite{McDonald:2003,Seljak:2012} and the HI velocity bias is set to one. We use the sign convention of these works and of ref.~\cite{Autieri:2026}. The \lya redshift-space distortion parameter is $\beta_{\rm F}=b_\eta f/b_{\rm F}>0$. The \meft adds to the Kaiser term the leading counterterms and stochastic terms:
\begin{align}
P_{\rm HI}(k,\mu)&=\Tb^2\Big[K_{\rm HI}^2P_{L}-2K_{\rm HI}\,c_{\rm HI}(\mu)\,k^2P_{L}+N_{00}^{\rm HI}+N_{20}^{\rm HI}\,k^2+N_{22}^{\rm HI}\,k^2\mu^2\Big],\label{eq:phh}\\
P_{\rm F}(k,\mu)&=K_{\rm F}^2P_{L}-2K_{\rm F}\,c_{\rm F}(\mu)\,k^2P_{L}+N_{00}^{\rm F}+N_{20}^{\rm F}\,k^2+N_{22}^{\rm F}\,k^2\mu^2,\label{eq:pff}\\
P_{\rm X}(k,\mu)&=\Tb\Big[K_{\rm HI}K_{\rm F}P_{L}-\big(K_{\rm F}\,c_{\rm HI}(\mu)+K_{\rm HI}\,c_{\rm F}(\mu)\big)k^2P_{L}+N_{00}^{\rm X}+N_{20}^{\rm X}\,k^2+N_{22}^{\rm X}\,k^2\mu^2\Big],\label{eq:phf}
\end{align}
with $c_a(\mu)=c_0^a+c_2^a\mu^2+c_4^a\mu^4$ ($a\in\{{\rm HI},{\rm F}\}$) in $h^{-2}{\rm Mpc}^2$, and the stochastic coefficients $N_{ij}$ of the terms $N_{ij}k^i\mu^j$ in $(h^{-1}{\rm Mpc})^{3+i}$ (\lref{Section}{sec:scales}).

Each spectrum has its own stochastic terms, all of the form $N_{00}+N_{20}k^2+N_{22}k^2\mu^2$. For the \lya auto-spectrum this is the form of ref.~\cite{Autieri:2026}, which maps onto the shot-noise and scale-dependent stochastic terms used there. The stochastic components of the two tracers can be correlated, since HI in halos and the surrounding gas probed by the \lya forest are affected by the same small-scale processes~\cite{Desjacques:2018}. The cross-spectrum therefore has its own noise terms, which are not fixed by those of the auto-spectra.

The model has three parameters that enter non-linearly, $\{\bHI,b_{\rm F},b_\eta\}$, and fifteen parameters that enter linearly: six counterterms and nine stochastic coefficients, which we marginalise analytically. Fitted to the data, the \meft has about 18 effective parameters, i.e.\ all its parameters are constrained by the data (\lref{Section}{sec:inference}).
We also tested adding the next-to-leading fingers-of-God counterterm $-\tilde c_{ab}\,f^4\mu^4k^4K_aK_bP_{L}$ to the 21\,cm auto- and cross-spectra ($(a,b)=({\rm HI},{\rm HI})$ and $({\rm HI},{\rm F})$)~\cite{Ivanov:2020,Chudaykin:2020}, and a 21\,cm noise term $N_{44}k^4\mu^4$ in place of $N_{20}k^2$, the leading expansion of the noise of a tracer whose small-scale velocities produce fingers of God~\cite{Perko:2016,Chen:2020}. Adding the counterterm lowers the joint $\chi^2$ at the reference cuts by 3--5 for two parameters, moves the linear biases by at most $0.36\sigma$ and increases their uncertainties by at most 3.2\%. The $N_{44}$ form of the 21\,cm noise increases $\chi^2$ by 1.1--9.2 and moves $\bHI$ by 0.9--$1.2\sigma$ (\lref{Section}{sec:robustness}). We include neither in the baseline \meft.

\begin{table}
\centering
\small
{\renewcommand{\arraystretch}{1.3}%
\begin{tabular}{lll}
\toprule
Parameter & Prior & Treatment \\
\midrule
$\bHI$; $b_{\rm F}$; $b_\eta$ & $\mathcal U[0,10]$; $\mathcal U[-2,0]$; $\mathcal U[-2,1]$ & sampled \\
\midrule
$c^{\rm HI}_{0,2,4}$; $c^{\rm F}_{0,2,4}$ [$h^{-2}{\rm Mpc}^2$] & $\mathcal N(0,10^2)$; $\mathcal N(0,1)$ & marginalised \\
$N^{\rm HI}_{00}$; $N^{\rm HI}_{20}$, $N^{\rm HI}_{22}$ & $\mathcal N(0,1000^2)$; $\mathcal N(0,5000^2)$ & marginalised \\
$N^{\rm X}_{00}$; $N^{\rm X}_{20}$, $N^{\rm X}_{22}$ & $\mathcal N(0,50^2)$; $\mathcal N(0,10^2)$ & marginalised \\
$N^{\rm F}_{00}$; $N^{\rm F}_{20}$, $N^{\rm F}_{22}$ & $\mathcal N(0,10^2)$; $\mathcal N(0,0.5^2)$ & marginalised \\
\bottomrule
\end{tabular}}
\caption{Parameters and priors of the \meft. The counterterms and the stochastic terms have the same priors at all redshifts. The stochastic coefficients $N_{ij}$ of the terms $N_{ij}k^i\mu^j$ are in $(h^{-1}{\rm Mpc})^{3+i}$. The counterterm widths are those of ref.~\cite{Chudaykin:2025} for halos (HI) and the \lya forest. The stochastic widths are 2.4--5 times the largest best-fit value over the five redshifts, for the \meft and for the one-loop EFT used to build the covariance (\lref{Section}{sec:covariance}), in fits with much wider priors, rounded, with the two $k^2$ coefficients of each spectrum sharing the width set by the larger of them. The parameters marked ``marginalised'' enter linearly and are integrated out analytically.}
\label{tab:priors}
\end{table}

\subsection{Relevant scales}
\label{sec:scales}

A perturbative model is expected to fail at the largest wavenumber at which a physical process not captured by the expansion becomes important. For the matter field this is the non-linear scale $k_{\rm NL}^{\rm DM}$, defined by $\Delta^2_{L}(k_{\rm NL}^{\rm DM})=k^3P_{L}/(2\pi^2)=1$. Biased tracers in redshift space have additional scales: the velocity dispersion of the objects that produce the signal, which generates fingers of God along the line of sight, and the spatial extent of the processes that set the tracer density. \lref{Table}{tab:scales} compares these scales.

The matter non-linear scale grows from $3.4$ to $30\hMpc$ between $z=2.4$ and $4.2$, well above the largest wavenumber we use. The HI fingers of God of \leqref{Eq.}{eq:fog} correspond to $1/\sigma_{\rm s}\simeq0.8$--$1.8\hMpc$ for halos of $10^{11}$--$10^{12}\,h^{-1}M_\odot$, almost independently of redshift. The \lya forest is smoothed by gas pressure on a scale $\lambda_{\rm P}\simeq80$\,kpc (proper) at $z\simeq2$~\cite{Gnedin:1998,Kulkarni:2015,Rorai:2017}, i.e.\ $1/\lambda_{\rm P}\simeq5.4$--$3.5\hMpc$ (comoving) from $z=2.4$ to 4.2, and by thermal broadening on $k_\parallel\sim10\hMpc$~\cite{Hui:1997}. Moreover, the linear displacement $\sigma_{\rm d}$ due to bulk flows decreases from $2.2$ to $1.5\Mpch$ over our redshift range. Its effect on the shape of the power spectrum is partly captured by the counterterms in the \meft.

These scales anticipate the ordering of the scale cuts (\lref{Section}{sec:cuts}). The 21\,cm auto-spectrum, which contains the HI fingers of God twice, should fail first. The cross-spectrum, which contains them once, should fail later, and the \lya auto-spectrum, limited by pressure smoothing, last.

The same scales could set the normalisation of the scale-dependent stochastic terms, if written in terms of $\kt=k/k_{\rm NL}^a$. In an EFT expansion each power of $k^2$ is suppressed by the scale at which the expansion breaks down, so that the dimensionless coefficients are of order unity. Since the stochastic terms enter linearly and are marginalised analytically, $k_{\rm NL}^a$ would only set the physical widths of their priors, which would scale as $(k_{\rm NL}^a)^{-i}$ for a $k^i$ term. We write instead both the counterterms and the stochastic terms in physical units, with the same priors at all redshifts (\lref{Table}{tab:priors}).

\begin{table}
\centering
\small
\begin{tabular}{lccccc}
\toprule
Scale [$h\,{\rm Mpc}^{-1}$] & $z=2.4$ & $2.8$ & $3.2$ & $3.6$ & $4.2$ \\
\midrule
Matter non-linear scale, $\Delta^2_{L}(k)=1$ & 3.4 & 5.5 & 8.7 & 14 & 30 \\
Linear displacement, $1/\sigma_{\rm d}$ & 0.45 & 0.50 & 0.55 & 0.60 & 0.68 \\
HI fingers of God, $1/\sigma_{\rm s}$, $M=10^{12}\,h^{-1}M_\odot$ & 0.76 & 0.75 & 0.74 & 0.73 & 0.71 \\
HI fingers of God, $1/\sigma_{\rm s}$, $M=10^{11}\,h^{-1}M_\odot$ & 1.83 & 1.82 & 1.80 & 1.79 & 1.76 \\
\lya pressure smoothing, $1/\lambda_{\rm P}$ & 5.4 & 4.9 & 4.4 & 4.0 & 3.5 \\
\lya thermal broadening ($T=1.5\times10^4$\,K) & 9.5 & 9.9 & 10.4 & 10.8 & 11.5 \\
\midrule
Largest non-linear scale, HI & 0.76 & 0.75 & 0.74 & 0.73 & 0.71 \\
Largest non-linear scale, \lya & 3.4 & 4.9 & 4.4 & 4.0 & 3.5 \\
Geometric mean (cross) & 1.6 & 1.9 & 1.8 & 1.7 & 1.6 \\
\bottomrule
\end{tabular}
\caption{Physical scales relevant to the validity of the model. $\sigma_{\rm s}$ is the line-of-sight HI velocity dispersion of \leqref{Eq.}{eq:fog} in comoving units, $\lambda_{\rm P}$ the comoving pressure-smoothing scale for a fixed proper scale of 80\,kpc (the actual scale depends on the thermal history), and the thermal broadening scale is $\sqrt2\,aH/b$ with $b=(2k_{\rm B}T/m_{\rm p})^{1/2}$. The last three rows give the largest non-linear scale of each spectrum, the smaller of $k_{\rm NL}^{\rm DM}$ and the tracer scale ($1/\sigma_{\rm s}$ for $10^{12}\,h^{-1}M_\odot$, $1/\lambda_{\rm P}$).}
\label{tab:scales}
\end{table}

\subsection{Multipoles, likelihood and inference}
\label{sec:inference}

The model is compared with the deconvolved multipoles of \lref{Section}{sec:multipoles}. Because the low-$k$ bins contain few modes, we evaluate the model on the individual modes $(|\mathbf k_i|,\mu_i)$ and apply the same matrix $W^{-1}$.

We use a Gaussian likelihood with the covariance of \lref{Section}{sec:covariance}, and since the model is linear in the counterterms and stochastic coefficients, we marginalise over these parameters analytically with their Gaussian priors. We sample the posterior distributions with the No-U-Turn sampler~\cite{Hoffman:2014} in \textsc{JAX}~\cite{JAX}. For the scans of \lref{Section}{sec:cuts} and the data combinations of \lref{Section}{sec:combinations} we use the maximum of the posterior and the Laplace approximation, which we validated against the full posteriors of the joint fits. 

To judge the goodness of fit with Gaussian priors on many parameters, we use the effective number of degrees of freedom $\nu_{\rm eff}=N_{\rm d}-{\rm tr}[F(F+\Pi)^{-1}]$~\cite{Raveri:2018wln}, where $N_{\rm d}$ is the number of data points, $F$ the Fisher matrix of all parameters at the best fit and $\Pi$ the prior precision matrix. A fit is acceptable if its $\chi^2$ lies below the 95th percentile of the $\chi^2$ distribution with $\nu_{\rm eff}$ degrees of freedom. At the reference cuts, ${\rm tr}[F(F+\Pi)^{-1}]\simeq18$ for the \meft.

\section{Scale cuts}
\label{sec:cuts}
\label{sec:criteria}

We determine the scale cuts empirically on the simulation, first for each spectrum alone and then for the joint fit, and we reduce the cut of the spectrum that limits the joint fit if it fails. The cut is the same for the three multipoles $\ell\le4$. At each cut we require that
\begin{enumerate}
\item the $\chi^2$ of the best fit lies below the 95th percentile of $\chi^2_{\nu_{\rm eff}}$ (\lref{Section}{sec:inference});
\item the linear biases do not move from their value at a conservative anchor cut by more than their uncertainty at the cut being tested (``$1\sigma$ drift'');
\item the same biases do not move by more than 2\% from their anchor values (``absolute drift'').
\end{enumerate}
The anchor cuts are $0.5\hMpc$ for the 21\,cm auto-spectrum, $0.8\hMpc$ for the cross-spectrum and $1.0\hMpc$ for the \lya auto-spectrum. The second criterion depends on how well the model constrains the biases. The third criterion sets a fixed tolerance on the theoretical error of the biases. We chose 2\% because it is comparable to the statistical uncertainty of the best-determined biases in the joint \meft fit (2.0--2.5\% for $\bHI$ and 2.1--2.3\% for $b_{\rm F}$, \lref{Table}{tab:biases}), so that a smaller shift would be hidden by the noise of the fit, and a larger one would be a significant systematic error at that precision. It is also larger than the scatter of the drift expected from noise alone, which for nested data sets is $\sigma_\Delta=(\sigma_{\rm anchor}^2-\sigma_{k}^2)^{1/2}$ and is below 0.5\% for $\bHI$ in the joint fit between 0.5 and $0.8\hMpc$ (\lref{appendix}{app:cuts}). \lref{Appendix}{app:cuts} describes the procedure and the scans in detail.

\lref{Table}{tab:cuts} summarises the resulting cuts, which are the same at all redshifts. The \lya auto-spectrum is fitted up to $2.8\hMpc$, but $|b_{\rm F}|$ drifts by 2--6\% between 1 and $2\hMpc$. In the joint fit, where $b_{\rm F}$ is set mostly by the combination with the other spectra, moving the \lya cut from 1.4 to $2.0\hMpc$ changes $b_{\rm F}$ by 0.7--2.5\% and $b_\eta$ by 2.8--3.7\% (at most $1.1\sigma$), and we adopt $\kmax^{\rm F}=2.0\hMpc$. The \meft fits the cross-spectrum with the \lya auto-spectrum up to $1.8\hMpc$. We adopt $1.2\hMpc$: beyond it, $b_{\rm F}$ and $b_\eta$ of this fit drift by 1.2--2.4\% already at $1.4\hMpc$ at $z\le3.2$, and the preference of the \meft for a $\Tb$ above the true value becomes more significant (\lref{Section}{sec:tb}). The \meft fits the 21\,cm auto-spectrum alone up to $1.8\hMpc$, but its $\bHI$ drifts by 7--9\% already at $0.6\hMpc$ (1.4--$1.8\sigma$) and by 10--13\% at $0.8\hMpc$, which sets its 21\,cm cut at $0.5\hMpc$. In the joint fit the 21\,cm cut could be extended to $0.8\hMpc$ at all redshifts, with $\chi^2$ below the 95th percentile and $\bHI$ within 1.7\% ($0.7\sigma$) of its value at $0.5\hMpc$, but we do not use in the joint analysis a cut at which the 21\,cm auto-spectrum alone fails.

Beyond these cuts the linear biases of the joint fit drift. Above the 21\,cm cut, $\bHI$ moves by at most 1.7\% at $0.8\hMpc$, and $b_\eta$ by up to 2.1\% at $z=4.2$. Above the cross cut the biases move by up to 2.1\%, and above the \lya cut by up to 8\% (\lref{appendix}{app:cuts}). The ordering of the cuts, \lya auto $>$ cross $>$ 21\,cm auto, matches the ordering of the scales of \lref{Table}{tab:scales}: the HI fingers of God enter the 21\,cm auto-spectrum twice and the cross-spectrum once, and the \lya forest is limited by the larger pressure-smoothing wavenumber.

\begin{table}
\centering
\small
\begin{tabular}{lcp{0.45\textwidth}}
\toprule
Spectrum & \meft & Criterion that sets the cut \\
\midrule
21\,cm auto, $\kmax^{\rm HI}$ & 0.5 & drift of $\bHI$ (21\,cm alone) \\
cross, $\kmax^{\rm X}$ & 1.2 & 2\% drift of $b_{\rm F}$, $b_\eta$ (X+F); $\Tb$ recovery (\lref{Section}{sec:tb}) \\
\lya auto, $\kmax^{\rm F}$ & 2.0 & drift of $b_{\rm F}$ (\lya alone); small effect in the joint fit \\
\bottomrule
\end{tabular}
\caption{Reference scale cuts in $h\,{\rm Mpc}^{-1}$, the same at all redshifts. At these cuts the \meft has 129 data points (15, 42 and 72 for the three spectra).}
\label{tab:cuts}
\end{table}

\section{Results of the minimal EFT}
\label{sec:meft}

\subsection{Goodness of fit and best-fit models}
\label{sec:bestfit}

At the reference cuts the joint \meft fit has $\chi^2=88.1$, $89.1$, $88.4$, $86.3$ and $80.2$ for 129 data points and $\nu_{\rm eff}\simeq111$ at $z=2.4$, $2.8$, $3.2$, $3.6$ and $4.2$, below the 95th percentile ($\simeq136.8$) at all redshifts. \lref{Figure}{fig:bestfit} compares the best fit with the data at $z=2.4$. The residuals of the three multipoles of the three spectra show no coherent trend with $k$. Two of the 129 residuals at $z=2.4$ (one or two at the other redshifts) exceed $2\sigma$ in absolute value, compared with about six expected for Gaussian fluctuations with an exact covariance.

\lref{Figure}{fig:terms} shows the contribution of each term to the monopole of the best fit. In the 21\,cm auto-spectrum the Kaiser term accounts for 74\% (64\%) of the model at the cut at $z=2.4$ ($4.2$), and the counterterms and noise terms make up the rest. In the \lya auto- and cross-spectra the counterterm and stochastic contributions at the cut have opposite signs and partly cancel: at $z=2.4$ they are $+0.78$ and $-0.85$ times the model for the \lya auto-spectrum and $+2.1$ and $-2.3$ for the cross-spectrum, with sums (dot-dashed lines) of only $-0.06$ and $-0.23$ times the model. In the \lya auto-spectrum at $z=4.2$ they reach $+3.9$ and $-3.7$ times the model, with a sum of $+0.21$, while in the cross-spectrum both are small ($-0.10$ and $-0.16$, with a sum of $-0.26$). The $k^2$ counterterms ($\propto k^2P_{L}$) and the $k^2$ stochastic terms ($\propto k^2$) differ only through the slope of $P_{L}$, which is close to $-2.3$ at $k\sim1$--$2\hMpc$, so their sum is much better determined than each term individually.

\begin{figure}
\centering
\includegraphics[width=\textwidth]{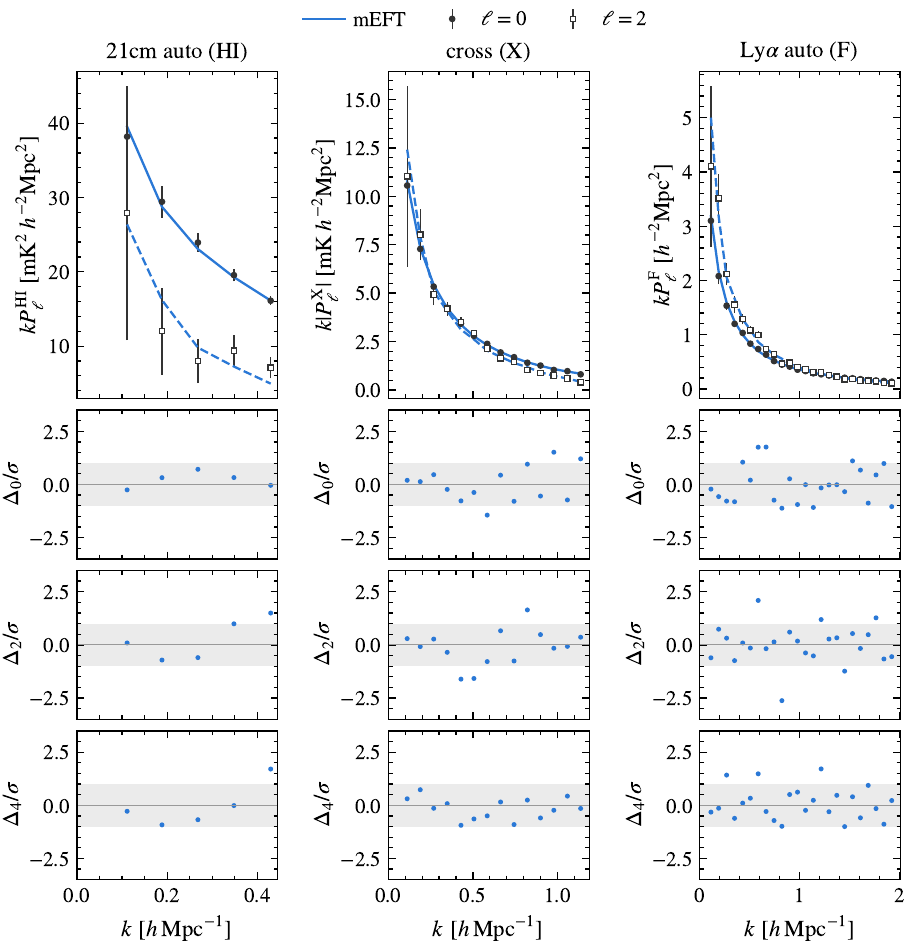}
\caption{Joint best fit of the \meft at $z=2.4$, at its reference cuts. Top: monopole (filled circles, solid lines) and quadrupole (open squares, dashed lines) of the three spectra, multiplied by $k$ (the cross-spectrum as its absolute value). Lower rows: residuals of the three multipoles in units of the uncertainty; the grey band marks $\pm1\sigma$.}
\label{fig:bestfit}
\end{figure}

\begin{figure}
\centering
\includegraphics[width=\textwidth]{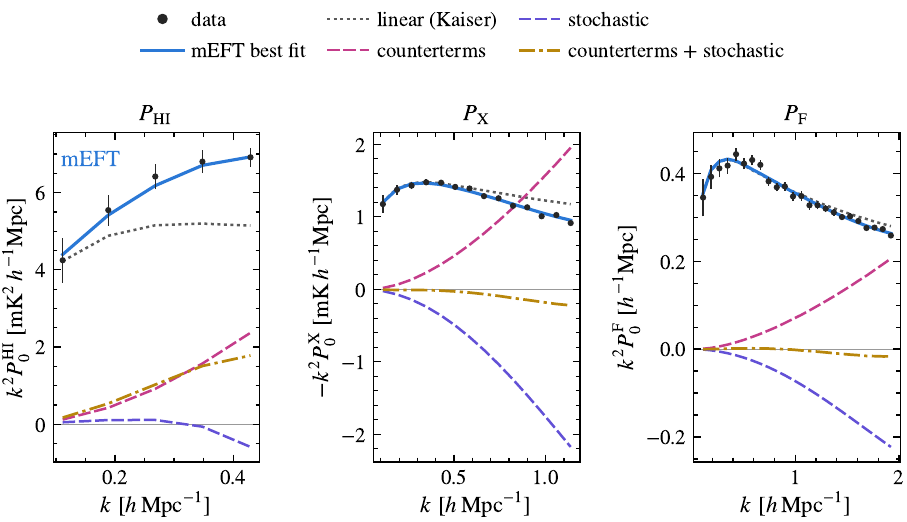}
\caption{Monopole of the joint best fit at $z=2.4$ against the data, multiplied by $k^2$ (the cross-spectrum with its sign flipped), for the \meft at its reference cuts, with the separate contributions: linear (Kaiser) term (dotted), counterterms and stochastic terms (dashed), and the sum of the counterterms and stochastic terms (dot-dashed), which partly cancel.}
\label{fig:terms}
\end{figure}

\subsection{What each data combination measures}
\label{sec:combinations}

The cross-spectrum alone constrains only the product $\Tb\bHI b_{\rm F}$ on large scales and $\Tb\bHI b_\eta$ through its anisotropy, so we always fit it together with at least one auto-spectrum. \lref{Figure}{fig:triangle} shows the posteriors of the linear biases from the 21\,cm auto-spectrum, the \lya auto-spectrum, the cross- and \lya auto-spectra (``X+F'') and the joint fit at $z=2.4$ and $4.2$, and \lref{Table}{tab:combos} lists all combinations. With $\Tb$ fixed, X+F determines $\bHI$ to 2.2--2.6\%, about three times as well as the 21\,cm auto-spectrum alone at its own cut (7.6--7.8\%), and the joint fit improves this to 2.0--2.5\%. The \lya biases are determined mostly by the \lya auto-spectrum, and adding the other two spectra reduces the uncertainties on $b_{\rm F}$ and $b_\eta$ by 2--10\%. The joint fit is therefore not much more precise than X+F for the biases, but it tests whether the three spectra are described consistently.

\begin{figure}
\centering
\includegraphics[width=0.5\textwidth]{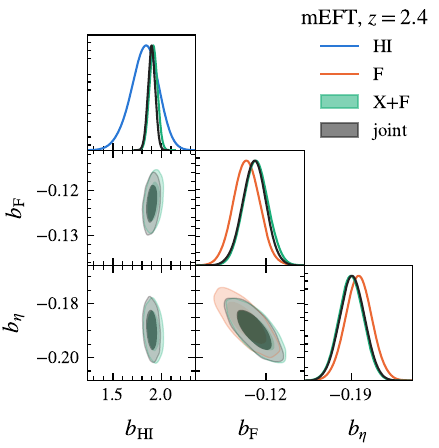}\hfill
\includegraphics[width=0.5\textwidth]{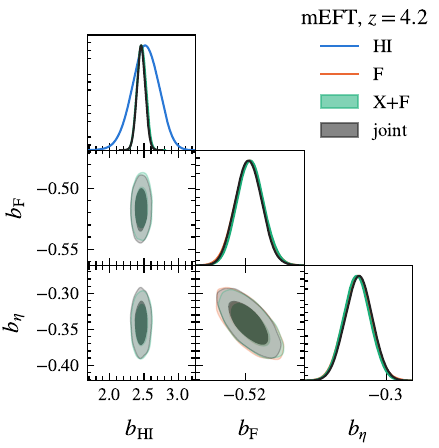}
\caption{Posteriors (68\% and 95\% contours) of the linear biases of the \meft from the 21\,cm auto-spectrum, the \lya auto-spectrum, the cross- and \lya auto-spectra, and the joint fit, at $z=2.4$ (left) and $4.2$ (right).}
\label{fig:triangle}
\end{figure}

\subsection{Consistency between data combinations}
\label{sec:consistency}

\lref{Figure}{fig:consistency} shows the linear biases from every combination of spectra relative to the joint fit. All combinations except one agree with the joint fit within about $0.6\sigma$. The combinations that include the 21\,cm auto-spectrum but not the cross-spectrum (HI and HI+F), whose $\bHI$ is three times less precise, give values within 4\% of the joint fit, lower at $z=2.4$ and higher at $z=4.2$. The exception is the combination of the cross- and 21\,cm auto-spectra (``X+HI''), which prefers $|b_\eta|$ higher by 13--18\% (1.5--$2.1\sigma$) and $|b_{\rm F}|$ lower by up to 7\% (about $1\sigma$). In X+HI the \lya kernel is set by the ratio of the cross-spectrum to the 21\,cm auto-spectrum. Therefore, at the precision of these data, the cross-spectrum prefers a more anisotropic \lya kernel than the \lya auto-spectrum, and the tree-level relation between the three spectra does not hold exactly. When the \lya auto-spectrum is included, $b_{\rm F}$ and $b_\eta$ are fixed by it and the difference is presumably absorbed by the counterterms and noise terms of the cross-spectrum.

\begin{figure}
\centering
\includegraphics[width=\textwidth]{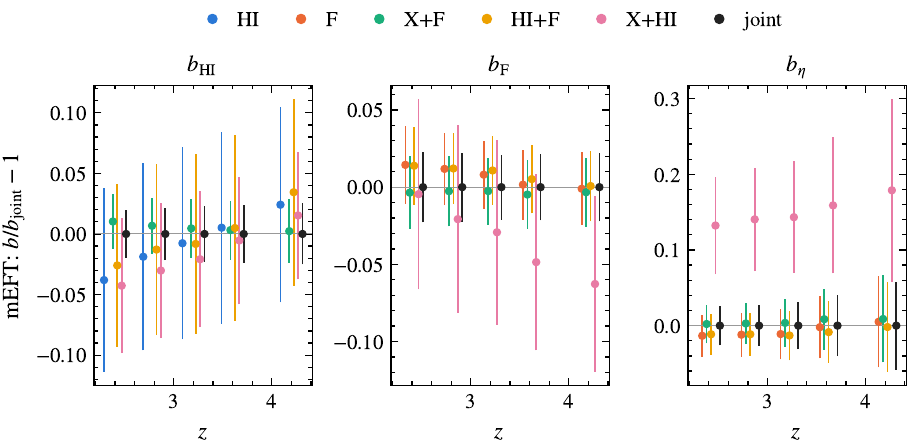}
\caption{Linear biases from each combination of spectra relative to the joint fit, for the \meft at the reference cuts. HI: 21\,cm auto; F: \lya auto; X: cross. Error bars are the uncertainties of the individual combinations.}
\label{fig:consistency}
\end{figure}

\begin{table}
\centering
\small
\begin{tabular}{cl ccc}
\toprule
$z$ & Data & $\bHI$ & $b_{\rm F}$ & $b_\eta$ \\
\midrule
\multirow{6}{*}{2.4} & HI & $1.841\pm0.139$ & -- & -- \\
 & F & -- & $-0.1246\pm0.0031$ & $-0.187\pm0.005$ \\
 & X+F & $1.920\pm0.042$ & $-0.1224\pm0.0029$ & $-0.190\pm0.005$ \\
 & HI+F & $1.853\pm0.127$ & $-0.1245\pm0.0030$ & $-0.187\pm0.005$ \\
 & X+HI & $1.829\pm0.103$ & $-0.1211\pm0.0074$ & $-0.215\pm0.012$ \\
 & joint & $1.897\pm0.038$ & $-0.1230\pm0.0028$ & $-0.189\pm0.005$ \\
\midrule
\multirow{6}{*}{3.2} & HI & $2.165\pm0.169$ & -- & -- \\
 & F & -- & $-0.2659\pm0.0059$ & $-0.311\pm0.010$ \\
 & X+F & $2.177\pm0.052$ & $-0.2632\pm0.0057$ & $-0.315\pm0.010$ \\
 & HI+F & $2.161\pm0.159$ & $-0.2665\pm0.0059$ & $-0.310\pm0.010$ \\
 & X+HI & $2.140\pm0.118$ & $-0.2530\pm0.0149$ & $-0.358\pm0.024$ \\
 & joint & $2.166\pm0.049$ & $-0.2639\pm0.0057$ & $-0.314\pm0.010$ \\
\midrule
\multirow{6}{*}{4.2} & HI & $2.545\pm0.197$ & -- & -- \\
 & F & -- & $-0.5169\pm0.0118$ & $-0.342\pm0.020$ \\
 & X+F & $2.463\pm0.064$ & $-0.5161\pm0.0116$ & $-0.343\pm0.020$ \\
 & HI+F & $2.557\pm0.186$ & $-0.5179\pm0.0117$ & $-0.340\pm0.020$ \\
 & X+HI & $2.509\pm0.129$ & $-0.4807\pm0.0286$ & $-0.404\pm0.042$ \\
 & joint & $2.459\pm0.062$ & $-0.5173\pm0.0117$ & $-0.341\pm0.020$ \\
\bottomrule
\end{tabular}
\caption{Linear biases from each combination of spectra at the reference cuts (maximum posterior and Laplace uncertainties, with $\Tb$ fixed), at $z=2.4$, $3.2$ and $4.2$. HI: 21\,cm auto-spectrum; F: \lya auto-spectrum; X: cross-spectrum.}
\label{tab:combos}
\end{table}

\subsection{Recovery of the brightness temperature}
\label{sec:tb}

So far we have fixed $\Tb$ to its true value. In a survey $\Tb$ is unknown, and a stated advantage of combining the 21\,cm and \lya data is that it can be measured. To test this, we multiply the 21\,cm auto-spectrum model by $a_{\rm T}^2$ and the cross-spectrum model by $a_{\rm T}$, where $a_{\rm T}\equiv\Tb/\Tb^{\rm true}$, and compute the profile likelihood on a grid of $a_{\rm T}$:
\begin{equation}
\chi^2_{\rm prof}(a_{\rm T})=\min_{\boldsymbol\theta}\,\min_{\boldsymbol\lambda}\,\chi^2(\boldsymbol\theta,\boldsymbol\lambda;a_{\rm T}),
\label{eq:chi2prof}
\end{equation}
where $\boldsymbol\theta$ are the linear biases and $\boldsymbol\lambda$ the counterterms and stochastic coefficients. Since the model is linear in $\boldsymbol\lambda$, the inner minimum is a generalised least-squares solution. By Wilks' theorem~\cite{Wilks:1938}, $\Delta\chi^2_{\rm prof}=1$ and 4 from the minimum give the 68\% and 95\% intervals. We use a profile instead of sampling $\Tb$ with the other parameters, as in recent profile-likelihood analyses of full-shape data~\cite{Herold:2022,Holm:2023}, because it tells directly how much the data disfavour the true $\Tb$ and is not affected by the projection effects~\cite{Simon:2023,Holm:2023} that the many nuisance parameters produce in marginal posteriors.

We do not include the Gaussian priors of $\boldsymbol\lambda$ in \leqref{Eq.}{eq:chi2prof}, and we do not marginalise over $\boldsymbol\lambda$, for the following reason. The HI counterterms and stochastic terms are coefficients of $\Tb\delta_{\rm HI}$, so their contributions to the model scale with $a_{\rm T}$ like the Kaiser term. With priors of fixed width on the coefficients, the range of amplitudes that these
terms can take in the measured spectra then shrinks as $a_{\rm T}$ decreases. The posterior used in all our other fits, with $\boldsymbol\lambda$ marginalised analytically,
\begin{equation}
-2\ln P(a_{\rm T})=\min_{\boldsymbol\theta}\Big[\min_{\boldsymbol\lambda}\big(\chi^2(\boldsymbol\theta,\boldsymbol\lambda;a_{\rm T})+\boldsymbol\lambda^{\rm T}\Pi_\lambda\boldsymbol\lambda\big)+\ln\det\big(F_\lambda+\Pi_\lambda\big)\Big]+{\rm const},
\label{eq:m2lnp}
\end{equation}
where $F_\lambda$ is the Fisher matrix of the linear parameters and $\Pi_\lambda$ their prior precision, depends on this range through the determinant, which favours low values of $a_{\rm T}$ irrespective of the data. For the 21\,cm and \lya auto-spectra together (``HI+F''), whose $\chi^2_{\rm prof}$ changes by at most 0.02 between the minimum and the true value, it places the minimum of $-2\ln P$ at $a_{\rm T}=0.37$--$0.51$. Dividing instead each coefficient by the power of $a_{\rm T}$ that multiplies it ($a_{\rm T}^2$ for the HI counterterms and 21\,cm noise, $a_{\rm T}$ for the \lya counterterms and the cross noise), so that its prior has a fixed width in the units of the measured spectra rather than per unit of $\Tb$, an equally legitimate choice of prior, changes the minimum of the joint fit from $a_{\rm T}=0.78$--$1.09$ to 1.25--1.35. Neither is preferred by the data, and the profile likelihood of \leqref{Eq.}{eq:chi2prof} does not depend on this choice.

\lref{Figure}{fig:tb} shows the profiles and \lref{Table}{tab:tb} summarises them. HI+F does not constrain $\Tb$: its profile is flat around the true value, and its 68\% interval spans about $\pm0.4$--0.7 at $z\le2.8$ and extends beyond our grid at higher redshift. In the \meft, $\Tb$ is separated from $\bHI$ only through the anisotropy of the HI spectra, and the 21\,cm auto-spectrum at $k\le0.5\hMpc$ does not determine it. The fits that include the cross-spectrum prefer a higher $\Tb$. The joint fit has its minimum at $a_{\rm T}=1.21$--$1.26$ and disfavours the true value by $\Delta\chi^2=3.5$, $4.1$, $3.1$, $2.4$ and $1.3$ at $z=2.4$--$4.2$, and X+F at $a_{\rm T}=1.20$--$1.30$ by $\Delta\chi^2=1.1$--4.9. The preference for a high $\Tb$ is present for all cross cuts from 0.8 to $1.6\hMpc$, with minima at $a_{\rm T}=1.14$--$1.45$ for the joint fit. Its significance grows with the cut, from $\Delta\chi^2\le1.2$ at $0.8\hMpc$ to 2.3--5.8 at $1.4\hMpc$ and 6.7--14.5 at $1.6\hMpc$, where the minimum also moves to higher $a_{\rm T}$, which is one more reason to keep the cross cut at $1.2\hMpc$.

With $\Tb$ free, the degeneracy between $\Tb$ and $\bHI$ is broken only by redshift-space distortions, and a higher $\Tb$ with a lower $\bHI$ changes both the anisotropy of the two HI spectra and the ratio of the cross- to the 21\,cm auto-spectrum. The fit appears to use this freedom to absorb the same mismatch between the tree-level cross- and auto-spectra that changes the \lya kernel preferred by X+HI (\lref{Section}{sec:consistency}). Adding the fingers-of-God counterterm $\tilde c_{ab}$ to the 21\,cm auto- and cross-spectra (\lref{Section}{sec:robustness}) reduces the tension only slightly: the joint fit then has its minimum at $a_{\rm T}=1.17$--$1.23$ and disfavours the true value by $\Delta\chi^2=0.8$--3.2, and X+F by 0.8--3.5. With $\Tb$ fixed, as in all other results of this paper, the effect is absorbed by the nuisance parameters at a small cost in $\chi^2$. However, a measurement of $\Tb$, and hence of $\Omega_{\rm HI}$, from the cross-spectrum with a tree-level model would be biased high by 20--30\%, at $1$--$2\sigma$ at the precision of these data.

\paragraph{Informative priors.} We repeated the profiles with Gaussian priors on $\Tb$ or on $\bHI$, added to $\chi^2_{\rm prof}$ (\lref{Figure}{fig:tbpriors}). At fixed cosmology $\Tb\propto\Omega_{\rm HI}h(1+z)^2/E(z)$, so a prior on $\Omega_{\rm HI}$ is a prior on $a_{\rm T}$ of the same fractional width. A prior on $\Tb$ pulls the result towards its centre but does not remove the bias: with widths of 10\% and 20\% the joint fit has its minimum at $a_{\rm T}=1.05$--1.10 and 1.11--1.18, and a 15\% prior centred 15\% above the true value gives 1.18--1.21. A prior on $\bHI$ breaks the degeneracy directly, because the amplitudes of the spectra then fix $\Tb$. A prior centred on the joint-fit $\bHI$ at the true $\Tb$, i.e.\ an ideal, unbiased calibration, gives $a_{\rm T}=1.01$, 1.02--1.03 and 1.05--1.10 for the joint fit with widths of 2\%, 5\% and 10\%, and similar values for X+F. As a realistic prior we computed $\bHI$ in the halo model, with the $M_{\rm HI}(M)$ relation of the painting (\lref{Section}{sec:painting}) and the mass function and halo bias of either refs.~\cite{Tinker:2008,Tinker:2010} (``T10'') or refs.~\cite{Sheth:1999,Sheth:2001} (``ST''), integrated with \textsc{Colossus}~\cite{Diemer:2018} over halos above the mass resolution of the catalogue. These values are 4--6\% (T10) and 8--10\% (ST) above the fitted $\bHI$, and the offset passes into $\Tb$: with a 5\% width the joint fit gives $a_{\rm T}=0.97$--1.00 (T10) and 0.95--0.97 (ST), and with a 10\% width 1.01--1.07 and 0.99--1.05. For HI+F the priors on $\bHI$ give $a_{\rm T}=0.81$--0.93, with 68\% intervals of $\pm0.15$--0.3 that include the true value, since the 21\,cm auto-spectrum at these scales constrains $\Tb\bHI$ only weakly. A $\Tb$ measurement with the \meft therefore requires an external calibration of $\bHI$, and is as accurate as that calibration.

\begin{figure}
\centering
\includegraphics[width=\textwidth]{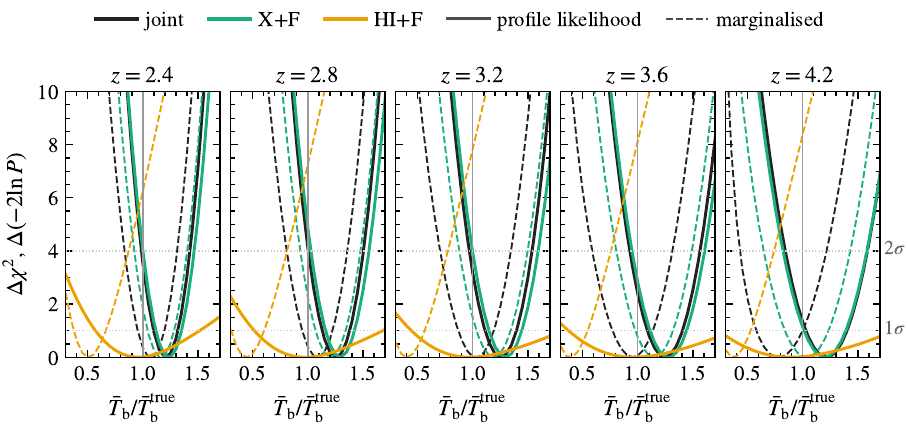}
\caption{Profile likelihood $\Delta\chi^2_{\rm prof}$ of \leqref{Eq.}{eq:chi2prof} (solid) as a function of $\Tb/\Tb^{\rm true}$ for the joint fit, the cross- and \lya auto-spectra (X+F) and the 21\,cm and \lya auto-spectra (HI+F), for the \meft at the five redshifts, on a grid of step 0.05. Dashed lines show $\Delta(-2\ln P)$ of \leqref{Eq.}{eq:m2lnp}, with the linear parameters marginalised and their priors of fixed width on the coefficients of the HI templates, which scale with $\Tb$. Dotted lines mark $\Delta=1$ and $4$, the 68\% and 95\% intervals of a single parameter.}
\label{fig:tb}
\end{figure}

\begin{table}
\centering
\small
\begin{tabular}{c ccc ccc ccc}
\toprule
 & \multicolumn{3}{c}{joint} & \multicolumn{3}{c}{X+F} & \multicolumn{3}{c}{HI+F} \\
$z$ & min & 68\% & $\Delta\chi^2$ & min & 68\% & $\Delta\chi^2$ & min & 68\% & $\Delta\chi^2$ \\
\midrule
2.4 & 1.21 & [1.10, 1.32] & 3.5 & 1.23 & [1.12, 1.35] & 4.0 & 0.96 & [0.57, 1.52] & 0.0 \\
2.8 & 1.26 & [1.13, 1.38] & 4.1 & 1.29 & [1.16, 1.42] & 4.9 & 0.98 & [0.51, 1.68] & 0.0 \\
3.2 & 1.25 & [1.11, 1.40] & 3.1 & 1.30 & [1.15, 1.44] & 3.9 & 0.98 & [0.44, $>1.8$] & 0.0 \\
3.6 & 1.25 & [1.09, 1.41] & 2.4 & 1.28 & [1.11, 1.44] & 2.8 & 0.96 & [0.37, $>1.8$] & 0.0 \\
4.2 & 1.21 & [1.03, 1.40] & 1.3 & 1.20 & [1.01, 1.39] & 1.1 & 0.89 & [$<0.3$, $>1.8$] & 0.0 \\
\bottomrule
\end{tabular}
\caption{Profile likelihood of $\Tb/\Tb^{\rm true}$ (\leqref{Eq.}{eq:chi2prof}), reporting the value at the minimum, 68\% interval ($\Delta\chi^2_{\rm prof}\le1$) and $\Delta\chi^2_{\rm prof}$ at the true value.}
\label{tab:tb}
\end{table}

\begin{figure}
\centering
\includegraphics[width=\textwidth]{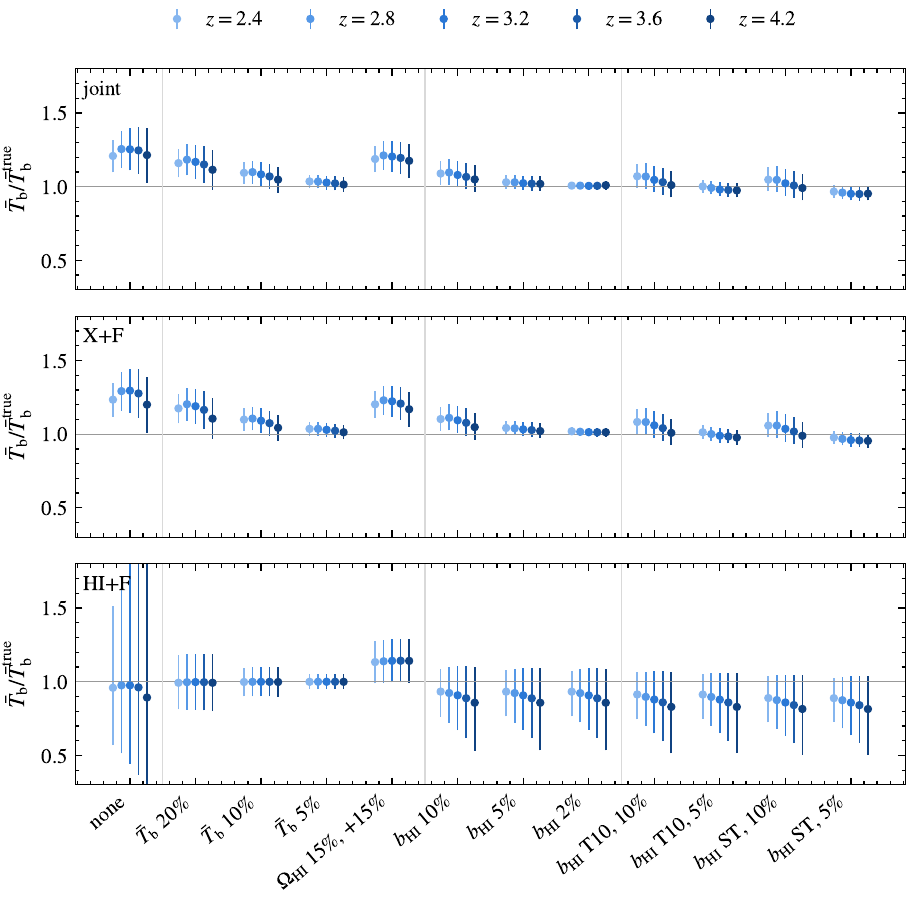}
\caption{$\Tb/\Tb^{\rm true}$ at the minimum of the profile likelihood, with the $\Delta\chi^2=1$ interval, without priors and with Gaussian priors on $\Tb$ (equivalently $\Omega_{\rm HI}$) or on $\bHI$: centred on the joint-fit value at the true $\Tb$ (``$\bHI$''), or on the halo-model values of refs.~\cite{Tinker:2008,Tinker:2010} (T10) and~\cite{Sheth:1999,Sheth:2001} (ST). Points at the five redshifts are offset horizontally.}
\label{fig:tbpriors}
\end{figure}

\subsection{Redshift evolution of the linear biases}
\label{sec:evolution}

\lref{Table}{tab:biases} lists the joint-fit biases at all redshifts. Between $z=2.4$ and $4.2$ the \meft gives $\bHI=1.899\pm0.038\to2.458\pm0.062$, $b_{\rm F}=-0.1228\pm0.0028\to-0.5171\pm0.0115$ and $b_\eta=-0.190\pm0.005\to-0.341\pm0.020$, so that $\beta_{\rm F}=b_\eta f/b_{\rm F}$ decreases from $1.50$ to $0.65$. All three evolve smoothly. The \lya biases from the \lya auto-spectrum alone at $z=2.4$, $b_{\rm F}=-0.1246\pm0.0031$ and $b_\eta=-0.187\pm0.005$ (\lref{Table}{tab:combos}), agree with those of ref.~\cite{Autieri:2026} for the same simulation. The increase of $|b_{\rm F}|$ and the decrease of $\beta_{\rm F}$ with redshift follow from the increase of the effective optical depth~\cite{ArinyoiPrats:2015}, as in ref.~\cite{Autieri:2026}. The increase of $\bHI$ reflects the larger relative bias of the halos that host HI at higher redshift~\cite{VillaescusaNavarro:2018,Castorina:2017}. A comparison with the true large-scale HI bias of the painted field would require its cross-spectrum with the matter field with the same variance reduction, which we leave for future work.

\begin{table}
\centering
\small
\resizebox{\textwidth}{!}{%
\begin{tabular}{cl cccccc}
\toprule
$z$ & Model (cuts) & $\bHI$ & $b_{\rm F}$ & $b_\eta$ & $\chi^2$ & $N_{\rm d}$ & $\nu_{\rm eff}$ \\
\midrule
2.4 & mEFT (0.5/1.2/2.0) & $1.899\pm0.038$ & $-0.1228\pm0.0028$ & $-0.190\pm0.005$ & 88.1 & 129 & 111.1 \\
2.8 & mEFT (0.5/1.2/2.0) & $2.039\pm0.044$ & $-0.1884\pm0.0042$ & $-0.267\pm0.007$ & 89.1 & 129 & 111.2 \\
3.2 & mEFT (0.5/1.2/2.0) & $2.166\pm0.050$ & $-0.2637\pm0.0056$ & $-0.314\pm0.010$ & 88.4 & 129 & 111.3 \\
3.6 & mEFT (0.5/1.2/2.0) & $2.278\pm0.055$ & $-0.3535\pm0.0076$ & $-0.333\pm0.013$ & 86.3 & 129 & 111.3 \\
4.2 & mEFT (0.5/1.2/2.0) & $2.458\pm0.062$ & $-0.5171\pm0.0115$ & $-0.341\pm0.020$ & 80.2 & 129 & 111.4 \\
\bottomrule
\end{tabular}}
\caption{Linear biases from the joint fits (posterior mean and standard deviation), with $\Tb$ fixed. Cuts are given as $\kmax^{\rm HI}/\kmax^{\rm X}/\kmax^{\rm F}$ in $h\,{\rm Mpc}^{-1}$. }
\label{tab:biases}
\end{table}

\subsection{Nuisance parameters}
\label{sec:nuisance}

\lref{Figure}{fig:nuisance} shows the best-fit counterterms and stochastic terms of the joint \meft fit, with their conditional uncertainties. The \lya counterterms are well determined and grow monotonically in absolute value with redshift: $c_0^{\rm F}$ from $0.044\pm0.020$ to $0.57\pm0.09\,h^{-2}{\rm Mpc}^2$ and $c_2^{\rm F}$ from $-0.072\pm0.023$ to $-0.46\pm0.06\,h^{-2}{\rm Mpc}^2$. The 21\,cm counterterms are less well determined at $\kmax^{\rm HI}=0.5\hMpc$ with $c_0^{\rm HI}$ growing from $-4.1\pm2.2$ to $3.2\pm4.3\,h^{-2}{\rm Mpc}^2$ and $c_2^{\rm HI}$ decreasing from $4.5\pm2.5$ to $-3.2\pm2.6\,h^{-2}{\rm Mpc}^2$. The constant 21\,cm noise $N_{00}^{\rm HI}$ grows from $51\pm83$ to $309\pm103\,h^{-3}{\rm Mpc}^3$, the isotropic $k^2$ term $N_{20}^{\rm HI}$ from $-490\pm160$ to $-680\pm180\,h^{-5}{\rm Mpc}^5$ (3.0--$3.7\sigma$ from zero), and $N_{22}^{\rm HI}$ is consistent with zero. The cross-spectrum noise terms differ from zero by up to $2.1\sigma$ ($N_{00}^{\rm X}$), $2.6\sigma$ ($N_{20}^{\rm X}$ from $-1.1\pm0.5$ at $z=2.4$ to $-2.9\pm1.3\,h^{-5}{\rm Mpc}^5$ at $z=3.6$) and $3.4\sigma$ ($N_{22}^{\rm X}=0.6$--$2.3\,h^{-5}{\rm Mpc}^5$). The counterterms do not decrease systematically with redshift, as they would if their natural size scaled with the matter non-linear scale (\lref{Section}{sec:scales}). The well-measured \lya counterterms increase by an order of magnitude. However, because of the degeneracy between counterterms and stochastic terms discussed in \lref{Section}{sec:bestfit}, we do not attempt a physical interpretation of individual coefficients. Empirical relations between them and the linear biases, of the kind derived in ref.~\cite{Autieri:2026} for the \lya forest, are a natural input for simulation-based priors and are left for future work.

\begin{figure}
\centering
\includegraphics[width=\textwidth]{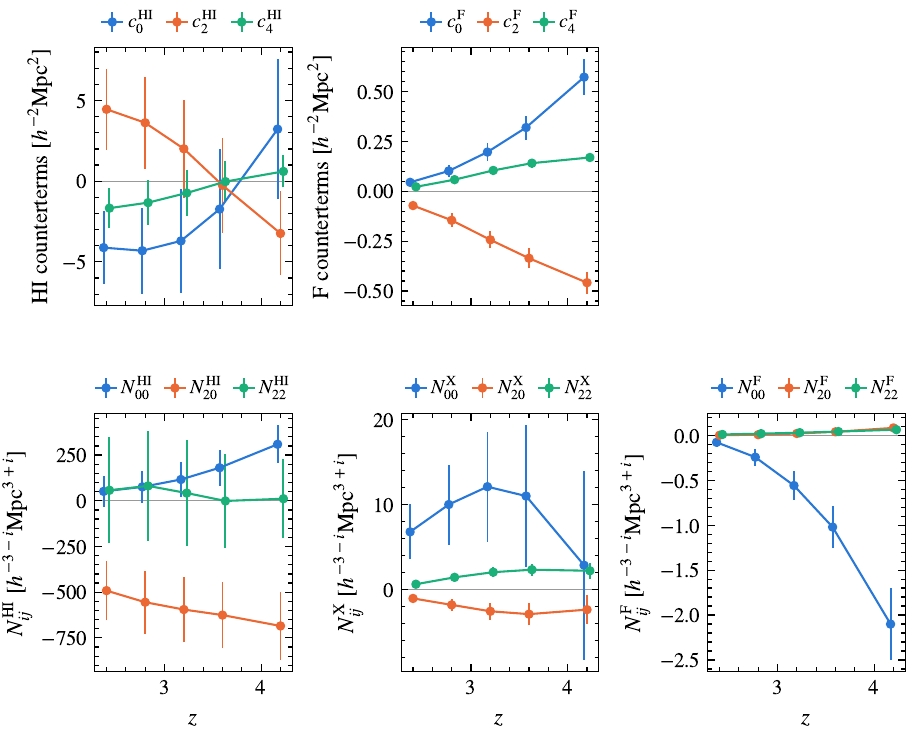}
\caption{Counterterms (top, in $h^{-2}{\rm Mpc}^2$) and stochastic coefficients $N_{ij}$ (bottom, in $h^{-3-i}{\rm Mpc}^{3+i}$) of the joint \meft fit as functions of redshift, with uncertainties conditional on the best-fit linear biases.}
\label{fig:nuisance}
\end{figure}

\subsection{Robustness}
\label{sec:robustness}

We tested the sensitivity of the \meft results to the priors and to the modelling choices. Widening or narrowing all counterterm and stochastic priors by factors between 0.5 and 10 moves the linear biases by at most $0.32\sigma$ (for halved priors at $z=4.2$; at most $0.20\sigma$ for wider priors), and changes the uncertainties by less than 8\%, confirming that the nuisance parameters are determined by the data, not by their priors.

We also tested the modelling choices of \lref{Section}{sec:meft_model} for the fingers-of-God counterterm and the stochastic terms with two additional sets of fits.
\begin{enumerate}
\item \meft with the fingers-of-God counterterm $-\tilde c_{ab}\,f^4\mu^4k^4K_aK_bP_{L}$ in the 21\,cm auto- and cross-spectra. It reduces the joint $\chi^2$ by 3--5 at the reference cuts, by 2--5 at $\kmax^{\rm HI}=0.6$--$0.7\hMpc$ and by 4--11 at $0.8\hMpc$, and moves the linear biases at the reference cuts by at most $0.36\sigma$. The drift of $\bHI$ in the joint fit changes by at most 0.4 percentage points up to $0.7\hMpc$ and 0.7 at $0.8\hMpc$. The drift of $\bHI$ in the 21\,cm auto-spectrum alone is reduced by only 1.5--2.4 percentage points at $0.6\hMpc$ (to 5.3--6.3\%) and 0.7--1.4 at $0.8\hMpc$ (to 9.6--11.5\%), so that with this term the 21\,cm cut of the \meft would remain at $0.5\hMpc$.
\item \meft with the 21\,cm noise $N_{00}+N_{22}k^2\mu^2+N_{44}k^4\mu^4$ in place of $N_{00}+N_{20}k^2+N_{22}k^2\mu^2$. $\chi^2$ increases by 1.1--9.2 and $\bHI$ moves by 0.9--$1.2\sigma$.
\end{enumerate}
Taken together, these tests show that the fingers-of-God counterterm improves the $\chi^2$ of the \meft but does not extend its 21\,cm cut, which is set by the drift of $\bHI$ in the 21\,cm auto-spectrum alone. At the reference cuts this counterterm moves the linear biases by at most $0.36\sigma$, while the $N_{44}$ noise form moves $\bHI$ by about $1\sigma$ and increases $\chi^2$. This justifies leaving both out of the baseline.
\section{Conclusions}
\label{sec:conclusions}

We have tested whether a minimal effective field theory of biased tracers in redshift space, consisting of the Kaiser model with the leading counterterms, shared between auto- and cross-spectra, and the leading stochastic terms, describes the 21\,cm auto-, \lya auto- and 21\,cm--\lya cross-power spectra of the Sherwood $80\Mpch$ simulation with HI painted onto halos, at $z=2.4$--$4.2$. We used data with sample variance reduced by Zel'dovich control variates and a hybrid Gaussian covariance, built from the best-fit model and the ensemble mean of the control field, that includes the covariance between the three spectra. Our main findings are the following.
\begin{enumerate}
\item Validated one spectrum at a time and then jointly, the \meft describes the three spectra with scale cuts $\kmax=0.5$, $1.2$ and $2.0\hMpc$ for the 21\,cm auto-, cross- and \lya auto-spectra at all five redshifts. The 21\,cm cut is set by the 21\,cm auto-spectrum alone, whose $\bHI$ drifts by 7--9\% between 0.5 and $0.6\hMpc$ and by 10--13\% up to $1.2\hMpc$. The joint fit alone would allow $0.8\hMpc$, but we do not use in the joint analysis a cut at which the 21\,cm auto-spectrum alone fails. Beyond the adopted cuts the linear biases drift systematically.
\item The \meft results do not depend on the priors of its nuisance parameters.
\item Adding a $k^4\mu^4$ fingers-of-God counterterm to the \meft does not extend its 21\,cm range and changes the linear biases at the reference cuts by at most $0.4\sigma$, while the $k^4\mu^4$ form of the 21\,cm noise moves $\bHI$ by up to $1.2\sigma$ and increases $\chi^2$, so we include neither in the baseline.
\item The ordering of the scale cuts, \lya auto $>$ cross $>$ 21\,cm auto, follows the HI velocity dispersion and the \lya pressure-smoothing scale, which depend little on redshift, rather than the matter non-linear scale, which grows from 3 to $30\hMpc$ over our redshift range. The same holds for the counterterms, which do not decrease with redshift as the matter non-linear scale would imply. We therefore write the counterterms and stochastic terms in physical units, with the same priors at all redshifts.
\item With $\Tb$ left free, $\Tb$ is degenerate with $\bHI$ and, through it, with the other model parameters. In the profile likelihood, which unlike the marginalised posterior does not depend on how the priors of the nuisance parameters are normalised, the fits that include the cross-spectrum settle at values 20--30\% above the true $\Tb$ (by up to $\Delta\chi^2=4.1$ for the joint fit and 4.9 for the cross- and \lya auto-spectra), as the free $\Tb$ absorbs small differences between the cross- and auto-spectra that are otherwise taken up by the nuisance parameters, while the two auto-spectra together do not constrain $\Tb$. Priors on $\Tb$ or $\Omega_{\rm HI}$ of realistic width do not remove the bias, whereas a prior on $\bHI$ of a few percent recovers $\Tb$, with any offset of the prior centre passing into $\Tb$: halo-model values of $\bHI$, 4--10\% above the fitted ones, give $\Tb$ 0--5\% low with a 5\% prior width. With $\Tb$ fixed, as in our baseline, the linear biases are unaffected.
\end{enumerate}
These results support the use of the \meft for the joint analyses of 21\,cm and \lya forest data, in the range of scales accessible to 21\,cm surveys. With few parameters, all of them determined by the data rather than by their priors, the \meft gives percent-level constraints on the linear biases from the three spectra together. Natural extensions are larger simulation volumes, other hydrodynamical models and HI assignments, and survey effects such as instrumental noise, foregrounds and the survey window. A comparison with the one-loop effective field theory of biased tracers on the same data, to establish where its additional parameters pay off, is left for future work.

\acknowledgments
MV and GA are supported by the INFN INDARK Grant. MV is also supported by the IDEAS SISSA grant.

 The simulations used in this work were performed using
the Joliot Curie supercomputer at the Tr\`es Grand Centre de Calcul (TGCC) and the DiRAC Data Intensive service (CSD3) at the
University of Cambridge, managed by the University of Cambridge
University Information Services on behalf of the STFC DiRAC HPC
facility (www.dirac.ac.uk). We acknowledge the Partnership for Advanced Computing in Europe (PRACE) for awarding us time on Joliot
Curie in the 16th call. The DiRAC component of CSD3 at Cambridge
was funded by BEIS, UKRI and STFC capital funding and STFC operations grants. DiRAC is part of the National e-Infrastructure. 
Post-processing was partly done on the Ulysses supercomputer at SISSA.

\textit{AI usage statement:} Anthropic's Claude (Opus 5.5) was used to review, refine and optimize human-written code, assist with bookkeeping and management of analysis runs and check the manuscript for grammatical errors. The authors have independently verified and validated all results and take full responsibility for the content of this paper.

\appendix
\section{Validation of the scale cuts}
\label{app:cuts}

This appendix describes the scans behind the cuts of \lref{Section}{sec:cuts}. For the 21\,cm and \lya auto-spectra we fit each spectrum alone. The cross-spectrum alone constrains only $\Tb\bHI b_{\rm F}$, so we fit it together with the \lya auto-spectrum at $\kmax^{\rm F}=2\hMpc$. The drifts quoted below are relative to the anchor cuts of \lref{Section}{sec:cuts} and cover all five redshifts unless stated otherwise. Where a drift is compared with the noise expected for nested data sets, we use $\sigma_\Delta=(\sigma_{\rm anchor}^2-\sigma_k^2)^{1/2}$.

\subsection{Single spectra}
\label{sec:single}

\lref{Figure}{fig:scanalone} shows the linear biases, relative to their values at the anchor cuts, and the goodness of fit of the spectra fitted alone as functions of $\kmax$ at $z=2.4$.

\begin{figure}[t]
\centering
\includegraphics[width=\textwidth]{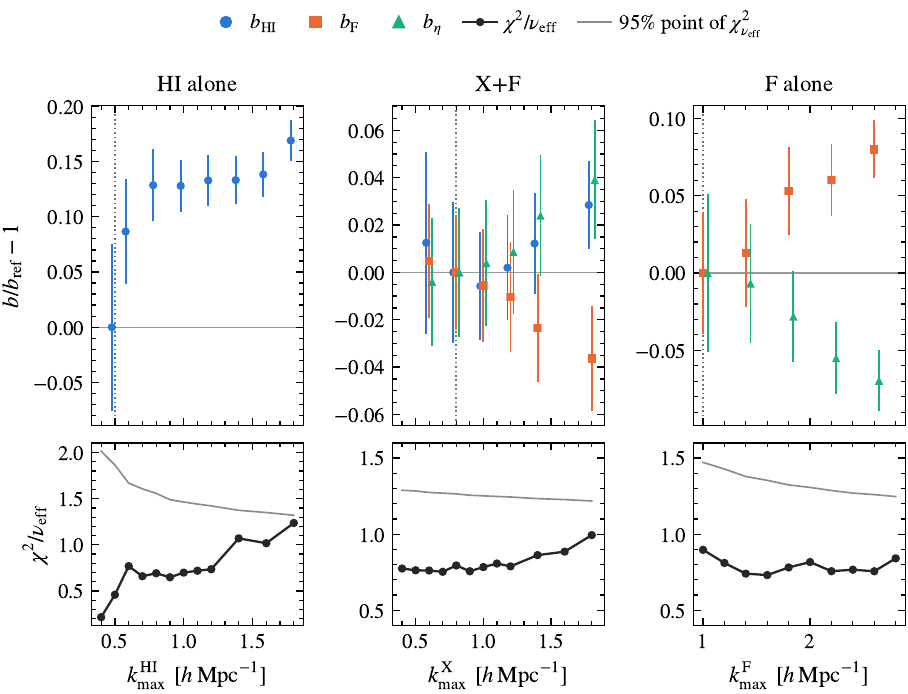}
\caption{Spectra fitted alone at $z=2.4$, as functions of the scale cut, for the \meft: 21\,cm auto-spectrum (left), cross-spectrum with the \lya auto-spectrum at $\kmax^{\rm F}=2\hMpc$ (centre) and \lya auto-spectrum (right). Top: linear biases relative to their value at the anchor cut of the drift criteria (\lref{Section}{sec:cuts}; dotted vertical lines), with their Laplace uncertainties; $\bHI$, $b_{\rm F}$ and $b_\eta$ are shown with different colours and markers (circles, squares and triangles). Points are offset horizontally for clarity, and only a subset of the cuts is shown. Bottom: $\chi^2/\nu_{\rm eff}$ at every cut, with the 95th percentile of $\chi^2_{\nu_{\rm eff}}/\nu_{\rm eff}$ (thin line).}
\label{fig:scanalone}
\end{figure}

\paragraph{\lya auto-spectrum.} The \meft fits the \lya auto-spectrum up to $\kmax=2.8\hMpc$, the largest cut we tested, at all redshifts, as in ref.~\cite{Autieri:2026}. However, $|b_{\rm F}|$ increases steadily with $\kmax$: by 1.1--1.3\% at $1.4\hMpc$, 2.1--5.8\% at $2.0\hMpc$ and 7--9\% at $2.8\hMpc$ relative to $\kmax=1\hMpc$, with a similar trend at all redshifts. Each of these steps moves $b_{\rm F}$ by less than about $2\sigma$. On its own, the absolute-drift criterion would place the \lya cut of the \meft at 1.4--$1.6\hMpc$ for $b_{\rm F}$, and at 1.0--$1.6\hMpc$ when $b_\eta$ is included. In the joint fit, however, $b_{\rm F}$ is set mostly by the combination with the other spectra: moving the \lya cut of the joint \meft fit from 1.4 to $2.0\hMpc$ changes $b_{\rm F}$ by 0.7--2.5\%, $b_\eta$ by 2.8--3.7\% (at most $1.1\sigma$) and $\bHI$ by at most 0.2\% (\lref{Figure}{fig:scanjoint}, right, at $z=2.4$). Relative to $\kmax^{\rm F}=1\hMpc$ the joint \meft $|b_{\rm F}|$ increases by 3.7\% ($z=2.4$) and 2.1\% ($z=4.2$) and $|b_\eta|$ decreases by 4.2\% and 4.0\%, at most $1.7\sigma$. These drifts exceed the 2\% criterion, and relative to $1\hMpc$ also the $1\sigma$ one at $z=2.4$, while $\bHI$ changes by at most 0.2\% between 1.4 and $2.0\hMpc$ (1.2\% relative to $1\hMpc$). We adopt $\kmax^{\rm F}=2.0\hMpc$.

\paragraph{21\,cm auto-spectrum.} The \meft fits the 21\,cm auto-spectrum alone up to $\kmax=1.8\hMpc$. Its $\bHI$ increases with the cut, by 7--9\% at $0.6\hMpc$, 10--13\% at $0.8\hMpc$ and 10--13\% at $1.2\hMpc$ relative to $\kmax=0.5\hMpc$, at every redshift. The uncertainty of $\bHI$ from the 21\,cm auto-spectrum alone decreases from 7.6--7.8\% at $0.5\hMpc$ to 2.0--2.2\% at $1.2\hMpc$, so the drift exceeds both $1\sigma$ and 2\% already at $0.6\hMpc$: both drift criteria place the 21\,cm cut of the \meft at $0.5\hMpc$ at every redshift, well below the $\chi^2$ limit.

\paragraph{Cross-spectrum.} With the \lya auto-spectrum at $2\hMpc$, the \meft fits the cross-spectrum up to $1.8\hMpc$. Relative to $\kmax^{\rm X}=0.8\hMpc$, $\bHI$ stays within 2\% up to 1.4--$1.8\hMpc$ at $z\le3.6$, while $b_{\rm F}$ and $b_\eta$ move by 1.2--2.4\% at $1.4\hMpc$ at $z\le3.2$. For the \meft at $z=4.2$, $\bHI$ shifts by $-2.2\%$ ($0.8\sigma$) at $1.0\hMpc$ and by $-2.1\%$ at $1.2\hMpc$, beyond the 2\% limit and close to the $1\sigma$ one.

\subsection{Joint fit}
\label{sec:joint_cuts}

\lref{Figure}{fig:scanjoint} shows the joint fit at $z=2.4$ when one cut is varied and the other two are kept at their reference values. We also ran the joint fit on a grid of 21\,cm and cross cuts, with the \lya cut at $2\hMpc$. With the 21\,cm cut at 0.5 or $0.6\hMpc$ the \meft joint fit is acceptable for cross cuts up to $1.6\hMpc$ at all redshifts, and at $0.6\hMpc$ its $\bHI$ moves by at most $0.6\%$ ($0.3\sigma$) relative to $0.5\hMpc$. With the 21\,cm cut at $0.8\hMpc$ it is acceptable for cross cuts up to $1.4\hMpc$ at all redshifts, while at $1.0\hMpc$ it exceeds the 95th percentile of $\chi^2$ for cross cuts from 0.9 to $1.1\hMpc$. The last 21\,cm shell of the adopted range ($k\simeq0.45\hMpc$) adds 2.5--4.2 to the joint $\chi^2$ for three data points, against 2.5--2.8 for the 21\,cm auto-spectrum alone, while the linear biases change by at most $0.4\sigma$.

\begin{figure}[t]
\centering
\includegraphics[width=\textwidth]{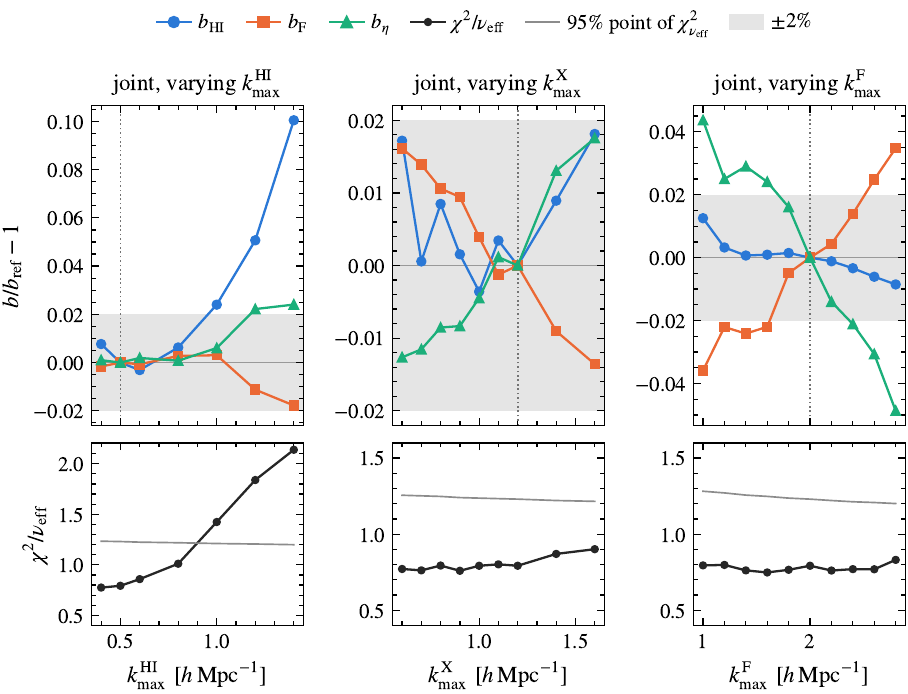}
\caption{Joint fit at $z=2.4$ as a function of one scale cut, with the other two at their reference values (0.5, 1.2 and $2.0\hMpc$ for the 21\,cm auto-, cross- and \lya auto-spectra). Left: 21\,cm cut; centre: cross cut; right: \lya cut. Same layout as \lref{Figure}{fig:scanalone}, but with the biases relative to their value at the reference cut (dotted vertical lines).}
\label{fig:scanjoint}
\end{figure}

The joint fit would therefore allow a 21\,cm cut of up to $0.8\hMpc$, larger than the one set by the drift criteria on the 21\,cm auto-spectrum alone but much smaller than the one allowed by its $\chi^2$ ($1.6\hMpc$). In the joint fit, sample variance largely cancels between the three spectra, and $\bHI$ is determined to 2.0--2.5\%, mostly by the cross- and \lya auto-spectra (2.2--2.6\% from these two alone), so that the 21\,cm modes above $0.5\hMpc$ add little information on $\bHI$ and the 10--13\% drift of the 21\,cm-only $\bHI$ between 0.5 and $1.2\hMpc$ does not propagate to the joint fit. We nevertheless keep $\kmax^{\rm HI}=0.5\hMpc$, since we do not use in the joint analysis a cut at which the 21\,cm auto-spectrum alone fails the drift criteria.

In the joint fit, the cross cut of the \meft is limited by $\bHI$ at high redshift: between $\kmax^{\rm X}=0.8$ and $1.2\hMpc$, $\bHI$ moves by $-2.5\%$ and $-2.7\%$ ($1.1\sigma$) at $z=3.6$ and $4.2$, against at most 1.2\% for $b_{\rm F}$ and $b_\eta$ at all redshifts. We keep $\kmax^{\rm X}=1.2\hMpc$ at all redshifts for uniformity, and note that it exceeds the limits of both drift criteria for the \meft at $z\ge3.6$.

\subsection{Beyond the adopted cuts}

\lref{Table}{tab:beyond} gives the drift of the joint-fit biases when one cut is increased beyond its reference value. Above the 21\,cm cut the \meft $\chi^2$ stays below the 95th percentile at 0.6 and $0.8\hMpc$, and $\bHI$ moves by at most 0.6\% at $0.6\hMpc$ and by $+0.3$ to $+1.7\%$ at $0.8\hMpc$. Here $\sigma_\Delta=(\sigma_{\rm ref}^2-\sigma_k^2)^{1/2}$ is the standard deviation of the drift expected from noise alone (\lref{Section}{sec:cuts}), with the reference cut in place of the anchor. Since the added 21\,cm modes carry little information on $\bHI$ in the joint fit, $\sigma_\Delta$ is at most 0.45\%. The drift at $0.8\hMpc$ is 1.2--$2.3\sigma_\Delta$ at $z\le3.6$, while at $z=4.2$, where the added modes do not reduce the uncertainty of $\bHI$ at all, its drift of 1.7\% is not consistent with noise. Above the cross cut, the biases move by up to 2.1\% at $1.6\hMpc$. Above the \lya cut, the \meft $|b_{\rm F}|$ increases by up to 3.5\% ($z=2.4$) and 4.5\% ($z=4.2$) at $2.8\hMpc$ and $|b_\eta|$ decreases by up to 4.9\% and 8.2\%.

\begin{table}[h!]
\centering
\small
\begin{tabular}{l c c ccc}
\toprule
cut & $\kmax$ & $z$ & $\bHI$ & $|b_{\rm F}|$ & $|b_\eta|$ \\
\midrule
\multirow{5}{*}{21\,cm} & \multirow{5}{*}{0.8} & 2.4 & $+0.6$ & $+0.3$ & $+0.1$ \\
 &  & 2.8 & $+0.6$ & $+0.7$ & $-0.3$ \\
 &  & 3.2 & $+0.3$ & $+0.9$ & $-0.5$ \\
 &  & 3.6 & $+0.5$ & $+1.0$ & $-0.8$ \\
 &  & 4.2 & $+1.7$ & $+1.0$ & $-2.1$ \\
\midrule
\multirow{5}{*}{cross} & \multirow{5}{*}{1.6} & 2.4 & $+1.8$ & $-1.4$ & $+1.8$ \\
 &  & 2.8 & $+1.9$ & $-1.5$ & $+2.1$ \\
 &  & 3.2 & $+1.4$ & $-1.0$ & $+2.0$ \\
 &  & 3.6 & $+0.9$ & $-0.6$ & $+1.8$ \\
 &  & 4.2 & $+0.2$ & $-0.3$ & $+1.6$ \\
\midrule
\multirow{2}{*}{\lya} & \multirow{2}{*}{2.8} & 2.4 & $-0.9$ & $+3.5$ & $-4.9$ \\
 &  & 4.2 & $-0.3$ & $+4.5$ & $-8.2$ \\
\bottomrule
\end{tabular}
\caption{Drift (in \%) of the joint-fit linear biases when one cut is increased beyond its reference value (0.5/1.2/2.0), with the other two at their reference values. The \lya cut was scanned at $z=2.4$ and $4.2$ only.}
\label{tab:beyond}
\end{table}

\bibliographystyle{JHEP}
\bibliography{references}

\end{document}